\documentclass[aps,prd,twocolumn, preprintnumbers, floatfix, superscriptaddress,showpacs,nofootinbib]{revtex4}
\usepackage[dvips]{graphics}
\usepackage{amsmath,amsfonts,bm}
\usepackage{epsfig,bm}
\usepackage{graphicx}
\def\MeijG[#1][#2][#3][#4][#5][#6]{G^{#1}_{#2}\left(#3,#4\left|\begin{array}{c}#5\\#6\end{array}\right|\right)}

\begin{document}

\preprint{LMU-ASC 66/12}

\affiliation{Dipartimento di Scienze Fisiche,
             Universit\'a di Napoli "Federico II", Via Cintia, 80126 Napoli, Italia}
\affiliation{INFN-Sezione di Napoli, Via Cintia, 80126 Napoli, Italia}
\affiliation{Ludwig-Maximilians-Universit\"at M\"unchen, Fakult\"at f\"ur Physik, Arnold Sommerfeld Center for Theoretical Physics, D–80333 M\"unchen, Germany}       

\author{Luigi Cappiello}
\affiliation{Dipartimento di Scienze Fisiche,
             Universit\'a di Napoli "Federico II", Via Cintia, 80126 Napoli, Italia}
\affiliation{INFN-Sezione di Napoli, Via Cintia, 80126 Napoli, Italia}

\author{Oscar Cat\`a}
\affiliation{Ludwig-Maximilians-Universit\"at M\"unchen, Fakult\"at f\"ur Physik, Arnold Sommerfeld Center for Theoretical Physics, D–80333 M\"unchen, Germany}       

\author{Giancarlo D'Ambrosio}
\affiliation{INFN-Sezione di Napoli, Via Cintia, 80126 Napoli, Italia}

\title{Standard Model prediction and new physics tests for $D^0\to h_1^+h_2^-{\ell}^+{\ell}^-(h=\pi,K; {\ell}=e,\mu)$}
\begin{abstract}
Motivated by the recent evidence for direct CP-violation in $D^0\to h^+h^-$ decays, we provide an exhaustive study of both Cabibbo-favored and Cabibbo-suppressed (singly and doubly) $D^0\to h_1^+h_2^-{\ell}^+{\ell}^-$ decays. In particular, we study the Dalitz plot for the long-distance contributions in the $(m_{ll}^2,m_{hh}^2)$ parameter space. We find that near-resonant effects, {\it{i.e.}}, $D^0\to V(h_1^+h_2^-){\ell}^+{\ell}^-$ with $V=\rho,K^*,\phi$, are sizeable and even dominant (over Bremsstrahlung) for the $\mu^+\mu^-$ decay modes, bringing the branching ratios close to the LHCb reach. We also provide a detailed study of the angular asymmetries for such decays and identify signatures for new physics detection. In particular, new physics signals can be neatly isolated in asymmetries involving the semileptonic operator $Q_{10}$, where for typical new physics scenarios the effects can be as sizeable as ${\cal{O}}(1\%)$ for the doubly Cabibbo-suppressed modes. 
\end{abstract}

\maketitle


\section{Introduction}\label{sec:I}
Processes involving flavor changing neutral currents (FCNC) are loop-suppressed in the standard model and therefore are especially suited as probes of new physics. In D decays the suppression is even more accentuated that in B or kaon decays: in charm physics FCNC involve down-type quarks and as a result the GIM mechanism is more efficient. Since new physics does not have to be subject to the same GIM suppression, charm decays are in principle an ideal arena to test physics beyond the standard model. The situation is of course not so simple: light quarks carry the bulk contribution to the decays, which means that they are long-distance dominated. The resulting hadronic uncertainties, due to their nonperturbative nature, are very difficult to estimate and overshadow short-distance effects, making their detection rather challenging. The situation gets worse because charm physics does not seem to accept an effective field theory description and one has to resort, for instance, to lattice computations or hadronic models. 

Quite recently the LHCb~\cite{Aaij:2011in} and CDF~\cite{Collaboration:2012qw} collaborations reported convincing evidence for direct CP violation in $D^0\to \pi^+\pi^-,K^+K^-$ decays. Specifically, they found that the CP asymmetry $\Delta a_{CP}\equiv a_{CP}^{KK}-a_{CP}^{\pi\pi}$ gives a nonzero value, that averaged with the previous results of B factories~\cite{Aubert:2007if,Staric:2008rx} gives~\cite{Amhis:2012bh}  
\begin{align}\label{CPexp}
\Delta a_{CP}=(-0.68\pm 0.15)\%
\end{align}
A standard model interpretation of the previous results is not ruled out~\cite{Bhattacharya:2012ah,Feldmann:2012js,Brod:2012ud} but seems hard to accomodate (see, {\it{e.g.}}~\cite{Isidori:2011qw,Franco:2012ck}), even allowing for generous uncertainties in the estimation of hadronic matrix elements. (This picture could change drastically, however, if the results of the latest LHCb analysis~\cite{Aaij:2013bra,Tilburg} are confirmed, where no significant deviation from the standard model is observed.)

If new physics is the explanation behind Eq.~(\ref{CPexp}), then one should scrutinize other decay modes in search of similar large effects. For instance, large CP-violating effects for radiative $D^0\to V\gamma$ decays would naturally point out at sizeable electromagnetic penguins~\cite{Isidori:2012yx}, which is a rather common feature of many extensions of the standard model.
 
In this paper we will study the semileptonic 4-body decays $D^0\to h_1^+h_2^-{\ell}^+{\ell}^-$, $({\ell}=e,\mu)$ in all its variants: Cabibbo allowed ($K^-\pi^+$), singly Cabibbo-suppressed ($\pi^+\pi^-; K^+K^-$) and doubly Cabibbo-suppressed ($K^+\pi^-$). Such a study is rather timely: LHCb has recently reported the potential for reaching branching ratios of $10^{-6}$ in the dimuon decays at the $3\sigma$ level~\cite{Bediaga:2012py}, improving previous upper bounds~\cite{Freyberger:1996it,Aitala:2000kk} by one order of magnitude. {\mbox{BESIII}}~\cite{BESIII} should also be able to study 4-body semileptonic $D^0$ decays in the near future.

On the theoretical side, there are a number of motivations to study these rare decays:
\begin{itemize}
\item The angular structure of a 4-body decay allows to define a variety of differential distributions. The associated Dalitz plots become essential tools for detailed tests of both the standard model and new physics. Of special importance are the different angular asymmetries that one can construct, which allow for a clean separation of short and long-distance effects.   
\item Access to short-distance physics is not limited to the charge asymmetry. In particular, some observables offer the possibility of disentangling new physics contributions in observables with tiny standard model backgrounds, like forward-backward asymmetries. This is especially interesting for the doubly Cabibbo-suppressed mode, which is particularly sensitive to new physics. 
\end{itemize}
 
In a nutshell, the penalty of small branching fractions one naturally pays in 4-body decays as opposed to 2 or 3-body decays is overly compensated by the diversity (and the size) of the asymmetries one can build. This implies that the semileptonic $D^0\to 2h2l$ decays have large potential to single out exceptionally clean experimental signatures.

The only reference for $D^0\to h_1^+h_2^-{\ell}^+{\ell}^-$ in the literature is the recent work in~\cite{Bigi:2011em}, where generic estimates are given for the branching ratios. In this work we will refine these estimates for the eight channels under study, detailing the differential distributions of the different long-distance contributions (bremsstrahlung and both magnetic and electric hadronic components) in the $(m_{\ell\ell}^2,m_{hh}^2)$ plane. Our results for the branching ratios are substantially larger than previously estimated in~\cite{Bigi:2011em}: $D^0\to K^+K^-{\ell}^+{\ell}^-$ hovers around $10^{-7}$, while we expect $D^0\to \pi^+\pi^-{\ell}^+{\ell}^-$ to be within the reach of LHCb. The Cabibbo-allowed mode $D^0\to K^-\pi^+{\ell}^+{\ell}^-$ is predicted in the high $10^{-6}$ while the doubly Cabibbo-suppressed $D^0\to K^+\pi^-{\ell}^+{\ell}^-$ is estimated at $10^{-8}$.

We will then proceed to study angular asymmetries that isolate short-distance effects. We will concentrate on two asymmetries that, due to tiny standard model backgrounds, are clean tests of new physics. First we will consider a T-odd asymmetry $A_{\phi}$, resulting from the interference between the electric and magnetic hadronic pieces, where $\phi$ is the angle between the dihadron and dilepton planes. This asymmetry involves four-quark operators and can be parametrized in terms of a weak phase $\delta_W$. Next we will consider the forward-backward asymmetry for the dilepton pair, $A_{FB}$, which involves the semileptonic penguin operator $Q_{10}$. While no generic prediction of their magnitude can be given (without resorting to models of new physics), we show that in both cases the signal is concentrated around the resonant region, {\it{i.e.}}, along the line defined by $m_{\ell\ell}^2\sim (0.5-1)$ GeV$^2$ and $m_{hh}^2=m_H^2$, where $H=\rho,K^{*},\phi$ for $hh=\pi^+\pi,K^-\pi^+,K^+K^-$, respectively. Taking some reference values for $\delta_W$ and $C_{10}$ we show that $A_{\phi}\sim (1-8)\%$, where the $D^0\to \pi^+\pi^-{\ell}^+{\ell}^-$ modes are the most favored, while $A_{FB}$ can reach ${\cal{O}}(1\%)$ for the doubly Cabibbo-suppressed decay $D^0\to K^+\pi^-\mu^+\mu^-$.   

We will organize this paper as follows: in Section~\ref{sec:II} we describe long distances, both Bremsstrahlung and hadronic contributions, and discuss their properties for the different decay modes in the ($m_{ll}^2, m_{hh}^2$) Dalitz plot. In Section~\ref{sec:IV} we first review the short-distance effects to $D^0\to 2h2{\ell}$ within the standard model and then discuss new physics scenarios that can enhance semileptonic operators while complying with $\Delta a_{CP}$ and current bounds from flavor physics. Section~\ref{sec:III} is devoted to angular asymmetries, where we study $A_{\phi}$ and $A_{FB}$ as two examples of clean tests of new physics. Conclusions are given in Section~\ref{sec:V}, while technical details are collected in three Appendices.


\section{Long-distance hadronic contributions}\label{sec:II}
\begin{figure}[t]
\begin{center}
\includegraphics[width=4.5cm]{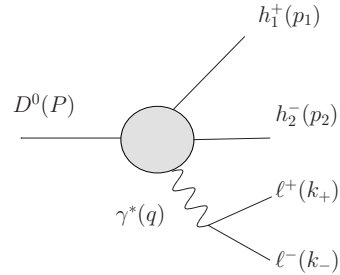}
\end{center}
\caption{\small{\it{Photon-mediated $D^0\to h_1^+h_2^-{\ell}^+{\ell}^-$ decay with our kinematic conventions. The blob represents the hadronic tensor $H_{\mu}$.}}}\label{fig:1}
\end{figure}

As mentioned in the Introduction, $D^0\to h_1^+h_2^-{\ell}^+{\ell}^-$ are largely dominated by long-distance effects. The bulk of the decay width thus comes from light down-type quarks running into one-loop diagrams. At low energies, when $\alpha_s\sim 1$ and dynamics become nonperturbative, the picture that applies is shown in Fig.~\ref{fig:1}, where the blob collects  hadronized strange and down quarks. At the same time, the dilepton pair creation is dominated by photon exchange. The amplitude for $D^0\to h_1^+h_2^-\gamma^*\to h_1^+h_2^-\ell^+\ell^-$ can be parametrized as 
\begin{align}\label{amplitude1}
{\cal M}_{LD}\equiv L^{\mu}(k_+,k_-) H_{\mu}(p_1,p_2,q)
\end{align} 
where $L^{\mu}$ is the leptonic current
\begin{align}
L^{\mu}(k_+,k_-)=\frac{e}{q^2} \big[\bar{u}(k_-)\gamma^\mu v(k_+)\big]
\end{align}
and $H_{\mu}$ is the hadronic vector, which can be written in terms of three form factors $F_{i}$:
\begin{equation}\label{param}
H^{\mu}(p_1,p_2,q)=F_1 p^\mu_1+F_2 p_2^\mu+F_3 \varepsilon^{\mu\nu\alpha\beta}p_{1\nu}p_{2\alpha}q_\beta
\end{equation}
The previous hadronic vector is also present in radiative $D^0\to h_1^+h_2^-\gamma$ decays. There it is common to perform a multipole expansion to distinguish the electric and magnetic components~\cite{D'Ambrosio:1992bf}, depending on whether the dihadron pair is in an intrinsic parity-even (electric) or parity-odd (magnetic) state. In terms of the form factors in Eq.~(\ref{param}), $F_{1,2}$ are electric contributions, while $F_3$ is the magnetic one. Here we will import this language to our decay.
   
In terms of the momenta $p_{1,2}$, $q=k_++k_-$ and $Q=k_+-k_-$, the squared amplitude can be cast as
\begin{align}\label{squared}
\sum_{\rm spins}|{\cal M}_{LD}|^2&=\frac{2e^2}{q^4}\Bigg[\sum_i^3|F_i|^2T_{ii}+2{\mathrm{Re}}\sum_{i<j}^3(F_i^{*}F_j)T_{ij}\Bigg]
\end{align}
where $(i,j=1,2)$
\begin{align}\label{Tkinematic}
T_{ij}&=(q\cdot p_i) (q\cdot p_j)-(Q\cdot p_i) (Q\cdot p_j)-q^2
(p_i\cdot p_j);\nonumber\\
T_{i3}&=(Q\cdot p_i)\epsilon_{\mu\nu\lambda\rho}p_1^{\mu}p_2^{\nu}q^{\lambda}Q^{\rho};\nonumber\\
T_{33}&=4 m_\ell^2\big[(m_{h1}^2m_{h2}^2-(p_1\cdot p_2)^2) q^2-m_{h1}^2(q\cdot p_2)^2\nonumber\\
&-m_{h2}^2(q\cdot p_1)^2+2(p_1\cdot p_2)(q\cdot p_1) (q\cdot p_2)\big]\nonumber\\
&+(Q\cdot p_1)^2\big[(q\cdot p_2)^2-q^2m_{h2}^2\big]\nonumber\\
&+(Q\cdot p_2)^2\big[(q\cdot p_1)^2-q^2m_{h1}^2\big]\nonumber\\
&+2(Q\cdot p_1)(Q\cdot p_2)\big[q^2 (p_1\cdot p_2)-(q\cdot p_1) (q\cdot p_2)\big]
\end{align}
Quite generally, there are three dynamically distinct long-distance contributions: (i) internal Bremsstrahlung, {\it{i.e.}}, QED radiation of photons away from the weak $D^0\to h_1^+h_2^-$ vertex; (ii) direct emission, where strong and weak effects get combined and resonance exchange is typically assumed to be the main contribution; and (iii) the charge radius contribution, where the photon gets radiated as a result of $D^0-{\overline{D^0}}$ mixing.

In the following subsections we will discuss, in turn, the Bremsstrahlung and direct emission contributions, giving expressions for the angular-integrated decay rates and branching ratios. The results will be shown in Dalitz plots for $d^2\Gamma/dq^2dp^2$.

In this paper we will not discuss the charge radius contribution since it bears no effect on the different angular asymmetries. Its contribution is limited to the decay width and subject to large uncertainties~\cite{Bigi:2011em}. With this in mind, our results for the decay width should thus be considered as a lower bound. However, it's hard to imagine that the charge radius can increase the decay width dramatically, so for the purposes of this paper omitting its contribution should be a reasonable approximation.   


\subsection{Bremsstrahlung}
The radiation of photons away from the weak $D^0\to h_1h_2$ vertex is a genuine infrared effect that can be computed with Low's theorem~\cite{Low:1958sn}:
\begin{align}\label{Brems}
{\cal{M}}_b(D^0\to h_1^+&h_2^-\gamma)=2e{\cal{M}}(D^0\to h_1^+h_2^-)\nonumber\\
&\times\left[\frac{p_1\cdot \epsilon}{2p_1\cdot q+q^2}-\frac{p_2\cdot \epsilon}{2p_2\cdot q+q^2}\right]
\end{align} 
Since the photon emission is factored out, the only ingredient that one needs is $|{\cal{M}}(D^0\to h_1^+h_2^-)|$, which can be extracted from experiment~\cite{PDG} for the different decay modes. Henceforth we will assume that there are no CP-violating phases in the $D^0\to h_1^+h_2^-$ amplitudes. 

In terms of the form factors entering $H_{\mu}$, Eq.~(\ref{Brems}) implies that 
\begin{align}\label{ib} 
F_j^{\rm (b)}&=(-1)^{j-1}\frac{2ie(1-\delta_{j3})}{2 q\cdot p_j+q^2} {\cal{M}}_{(D\to h_1h_2)}
\end{align}

Inserting the previous expression in Eq.~(\ref{squared}), one can easily build the differential decay rate. The kinematics of 4-body decays requires five variables. In this paper we will use the Cabibbo-Maksymowicz set of variables, $X_i=(q^2,p^2,\theta_h,\theta_{\ell},\phi)$, whose concrete definitions are given in Appendix~\ref{sec:VI}. In terms of these variables, one can express the differential decay rate quite generically as~\cite{Cabibbo:1965zz,Pais:1968zz}
\begin{align}\label{angular}
\frac{d^5\Gamma}{dxdy}&={\cal{A}}_1(x)+{\cal{A}}_2(x)s_{\ell}^2+{\cal{A}}_3(x)s_{\ell}^2c_{\phi}^2+{\cal{A}}_4(x) s_{2{\ell}}c_{\phi}\nonumber\\
&+{\cal{A}}_5(x) s_{\ell}c_{\phi}+{\cal{A}}_6(x)c_{\ell}+{\cal{A}}_7(x) s_{\ell}s_{\phi}\nonumber\\
&+{\cal{A}}_8(x)s_{2\ell}s_{\phi}+{\cal{A}}_9(x)s_{\ell}^2s_{2\phi}
\end{align}
where the vectors $x=(q^2,p^2,\cos{\theta_h})$ and $y=(\cos{\theta_{\ell}},\phi)$ split the dynamical variables (entering the form factors ${\cal{A}}_i(x)$) from the pure kinematical angular distribution described by $y$. In the following we will integrate the full angular dependence above and concentrate on the quantity $d^2\Gamma/dq^2dp^2$. Details thereof, including the analytical formulae, can be found in Appendix~\ref{sec:VIII}.

In Fig.~\ref{fig:3} we show the Dalitz plot in the $(q^2,p^2)$ plane for the different decay modes. Since Bremsstrahlung is an infrared effect driven by photon emission, one expects a significant contribution only in the low-$q^2$ region, with a sharp increase close to the dilepton threshold, whose maximum is reached in the region of large hadron recoil ($p^2\sim (m_D-2m_{\ell})^2)$. In Table~\ref{tab1} one can read off the resulting branching ratios for the different decay modes. For dimuon decays, branching ratios span between $10^{-7}-10^{-10}$, where the differences are solely due to the hierarchy between the hadronic branching ratios Br($D^0\to h_1h_2$). For electron-positron decays there is an increase of two orders of magnitude per channel, with branching ratios in the window $10^{-5}-10^{-8}$. This increase precisely illustrates the strength of Bremsstrahlung at low-$q^2$: electron-positron decays have a lower $q^2$-threshold and therefore probe infrared physics deeper. Notice that our results for the Bremsstrahlung are substantially larger than the rough estimates given in~\cite{Bigi:2011em} by 2-3 orders of magnitude.

Finally, we want to emphasize that the previous results for the Bremsstrahlung contribution only rely on Low's theorem ({\it{i.e.}}, QED) and experimental input for $D^0\to h_1h_2$. Therefore, the results in the first column of Table~\ref{tab1} should be seen as a solid lower estimate of the total branching ratios for the different decay channels. 
    
\begin{table*}[t]
\begin{center}
\begin{tabular}{|cccc|}
\hline
Decay mode &\,\,\,\, Bremsstrahlung &\,\,\,\, Direct emission (E) &\,\,\,\, Direct emission (M)\\
\hline
$D^0\to K^-\pi^+e^+e^-$ &\,\,\,\, $9.9\cdot 10^{-6}$ &\,\,\,\, $6.2\cdot 10^{-6}$ &\,\,\,\, $4.8\cdot 10^{-7}$\\
$D^0\to \pi^+\pi^-e^+e^-$ &\,\,\,\, $5.3\cdot 10^{-7}$ &\,\,\,\, $1.3\cdot 10^{-6}$ &\,\,\,\, $1.3\cdot 10^{-7}$\\
$D^0\to K^+K^-e^+e^-$ &\,\,\,\, $5.4\cdot 10^{-7}$ &\,\,\,\, $1.1\cdot 10^{-7}$ &\,\,\,\, $5.0\cdot 10^{-9}$\\
$D^0\to K^+\pi^-e^+e^-$ &\,\,\,\, $3.7\cdot 10^{-8}$ &\,\,\,\, $1.7\cdot 10^{-8}$ &\,\,\,\, $1.3\cdot 10^{-9}$\\
\hline
$D^0\to K^-\pi^+\mu^+\mu^-$ &\,\,\,\, $8.6\cdot 10^{-8}$ &\,\,\,\, $6.2\cdot 10^{-6}$ &\,\,\,\, $4.8\cdot 10^{-7}$\\
$D^0\to \pi^+\pi^-\mu^+\mu^-$ &\,\,\,\, $5.6\cdot 10^{-9}$ &\,\,\,\, $1.3\cdot 10^{-6}$ &\,\,\,\, $1.3\cdot 10^{-7}$\\
$D^0\to K^+K^-\mu^+\mu^-$ &\,\,\,\, $3.3\cdot 10^{-9}$ &\,\,\,\, $1.1\cdot 10^{-7}$ &\,\,\,\, $5.0\cdot 10^{-9}$\\
$D^0\to K^+\pi^-\mu^+\mu^-$ &\,\,\,\, $3.3\cdot 10^{-10}$ &\,\,\,\, $1.7\cdot 10^{-8}$ &\,\,\,\, $1.3\cdot 10^{-9}$\\
\hline
\end{tabular}
\end{center}
\caption{\small{\it{Long-distance contributions to the branching ratio for the different decay modes.}}}\label{tab1}
\end{table*}


\subsection{Resonant contributions}
Besides Bremsstrahlung, there are also long-distance contributions associated with hadronic effects, the most important of which are the near-resonant regions in both the dihadron and dilepton sectors. Depending on the strength of the resonant effect, these contributions have the potential of overcoming the Bremsstrahlung contribution discussed in the previous Section. Similar contributions have been studied in the $B\to K^* (K \pi) l^+l^-$ decay~\cite{Altmannshofer:2008dz}. The situation is in sharp contrast with what one encounters in kaon 4-body decays~\cite{Elwood:1995xv,Pichl:2000ab,Cappiello:2011qc} at least in two aspects: (i) in $Ke4$ decays the energy range is always far below resonance thresholds. As a result, Bremsstrahlung is overwhelmingly dominant; and (ii) additionally, the kaon direct emission contribution can be estimated in chiral perturbation theory. 

In the present case, the phase space is larger and decays like $D^0\to V^0 {\ell}^+{\ell}^-$ have nonnegligible branching ratios~\cite{Burdman:1995te}. As a result, $D^0\to V^0(h_1^+h_2^-){\ell}^+{\ell}^-$ are expected to play a significant role. However, unlike kaon decays, such contributions cannot be estimated with effective field theories and one has to resort to hadronic models. Therefore, one has to exercise some caution when interpreting the results: hadronic effects are genuinely nonperturbative and as such uncertainties are difficult to estimate. One should bear in mind that the results of this Section are not on the same solid ground as the Bremsstrahlung contributions discussed above.
     
\begin{figure}[t]
\begin{center}
\includegraphics[width=6.5cm]{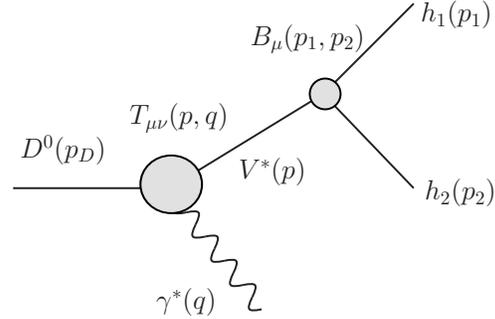}
\end{center}
\caption{\small{\it{Diagrammatic representation of the hadronic model adopted, with the weak $T_{\mu\nu}(p,q)$ and strong $B_{\mu}(p_1,p_2)$ correlators.}}}\label{fig:2}
\end{figure}

Quite generally, we will be considering underlying processes of the form depicted in Fig.~\ref{fig:2} 
where 
\begin{align}
T^V_{\mu\nu}(p,q)&=i^3\!\!\!\int d^4x e^{(ip\cdot x+iq\cdot y)}\langle 0| J^{{\bar{u}}c}(0)J_{\mu}^{V}(x)J_{\nu}^{\gamma}(y)|0\rangle\nonumber\\
&=t_1^Vq_{\mu}q_{\nu}+t_2^Vq_{\mu}p_{\nu}+t_3^Vg_{\mu\nu}+t_4^V\epsilon_{\mu\nu\lambda\rho}p_{\lambda}q_{\rho}\nonumber\\
&+t_5^Vp_{\mu}p_{\nu}+t_6^Vp_{\mu}q_{\nu}
\end{align}
and
\begin{align}
B^V_{\alpha}(p_1,p_2)&=i^3\!\!\!\int \!\! d^4x e^{(ip_1\cdot x+ip_2\cdot y)}\langle 0| J_{\alpha}^{V\dagger}(0)J_{1}(x)J_{2}(y)|0\rangle\nonumber\\
&=b_1^Vp_{1\alpha}-b_2^Vp_{2\alpha}
\end{align}
parameterize the most general tensorial decomposition of the $D^0\to V(p) \gamma(q)$ and $V(p)\to h_1(p_1) h_2(p_2)$ vertices. $t_i^V$ and $b_i^V$ are functions of the kinematical invariants of the problem, namely $t_i^V(q^2,p^2)$ and $b_j^V(p^2)$, where the upper index denotes their dependence on the exchanged resonance. 

The hadronic vector $H_{\mu}$ can be expressed as
\begin{align}
H_{\mu}(p_1,p_2,q)&=\sum_V\langle h_1h_2|{\cal{H}}|V\rangle\frac{\epsilon_{\mu}^{(\gamma)}(q)}{P_V(p^2)}\langle V\gamma^*|{\cal{H}}|D^0\rangle
\end{align}
where we will describe particle widths with a Breit-Wigner propagator, {\it{i.e.}}, $P_j(q^2)=q^2-m_j^2+i\Gamma_jm_j$, and
\begin{align}\label{formfactors}
\langle h_1 h_2| {\cal{H}} | V \rangle&=B_V^{\mu}(p_1,p_2)\epsilon_{\mu}^{(V)}(p)\nonumber\\
\langle V \gamma^*| {\cal{H}} | D^0\rangle&=T_V^{\mu\nu}(p,q)\epsilon_{\mu}^{(V)*}(p)\epsilon_{\nu}^{(\gamma)*}(q)
\end{align} 
In order to evaluate the matrix elements above we need to determine the functions $b_i$ and $t_i$. The former can be easily determined from the experimental $V\to h_1h_2$ decays. Notice that for the equal mass case, charge conjugation invariance (in our case $V$ is neutral) imposes that $b_1^V=b_2^V$. When $m_{h_1}\neq m_{h_2}$ the form factors differ by a term proportional to the mass difference (see Appendix~\ref{sec:VII}). In general, it is a good approximation to neglect this effect and henceforth we will use that $b_1^V=b_2^V\equiv b_V$ for all decay modes. We will also assume that the momentum dependence is soft and that to a good approximation $b_V(p^2)=b_V(m_V^2)\equiv b_V$.

The $V\to h_1h_2$ decay width is
\begin{align}
\Gamma_{V\to h_1h_2}&=\frac{1}{48\pi}b_V^2 m_V^5\lambda^{3/2}(m_V^2,m_{h1}^2,m_{h2}^2)
\end{align}
and comparison with the experimental determinations~\cite{PDG} gives
\begin{align}\label{Vhh}
b_{\rho}=5.92\,\, {\mathrm{GeV}};\,\,\,\, b_{K^*}=5.46\,\, {\mathrm{GeV}};\,\,\,\,b_{\phi}=4.41\,\, {\mathrm{GeV}}
\end{align}
The contribution of the $\omega(782)$ to $D^0\to \pi^+\pi^-{\ell}^+{\ell}^-$ is extremely suppressed ($b_{\omega}/b_{\rho}\simeq 3\%$) and will be neglected.

The evaluation of the weak vertex requires the $\Delta c=1$ effective hamiltonian. For the decay channels that we are considering, the relevant operators for the Cabibbo-allowed (C), singly Cabibbo-suppressed (SCS) and doubly Cabibbo-suppressed (DCS) transitions are 
\begin{align}
{\cal{H}}_{\Delta c=1}^{C}&=\frac{G_F}{\sqrt{2}}\left[\lambda_{sd} C_2^{(sd)}Q_{sd}\right]\nonumber\\
{\cal{H}}_{\Delta c=1}^{SCS}&=\frac{G_F}{\sqrt{2}}\left[\lambda_d C_2^{(d)} Q_d+\lambda_s C_2^{(s)} Q_s\right]\nonumber\\
{\cal{H}}_{\Delta c=1}^{DCS}&=\frac{G_F}{\sqrt{2}}\left[\lambda_{ds} C_2^{(ds)}Q_{ds}\right]
\end{align} 
where $\lambda_j=V_{cj}^*V_{uj}$, $\lambda_{sd}=V_{cs}^*V_{ud}$, $\lambda_{ds}=V_{cd}^*V_{us}$ and
\begin{align}
Q_{sd}&=({\bar{u}}\gamma_{\mu}c)_L({\bar{s}}\gamma^{\mu}d)_L\nonumber\\
Q_d&=({\bar{u}}\gamma_{\mu}c)_L({\bar{d}}\gamma^{\mu}d)_L\nonumber\\
Q_s&=({\bar{u}}\gamma_{\mu}c)_L({\bar{s}}\gamma^{\mu}s)_L\nonumber\\
Q_{ds}&=({\bar{u}}\gamma_{\mu}c)_L({\bar{d}}\gamma^{\mu}s)_L
\end{align}
In the previous equations we have neglected the penguin operators. To proceed further one needs to make some approximations. In this work we will assume that the factorizable contributions are the dominant ones, which means that 
\begin{figure}[h!]
\begin{center}
\includegraphics[width=8.5cm]{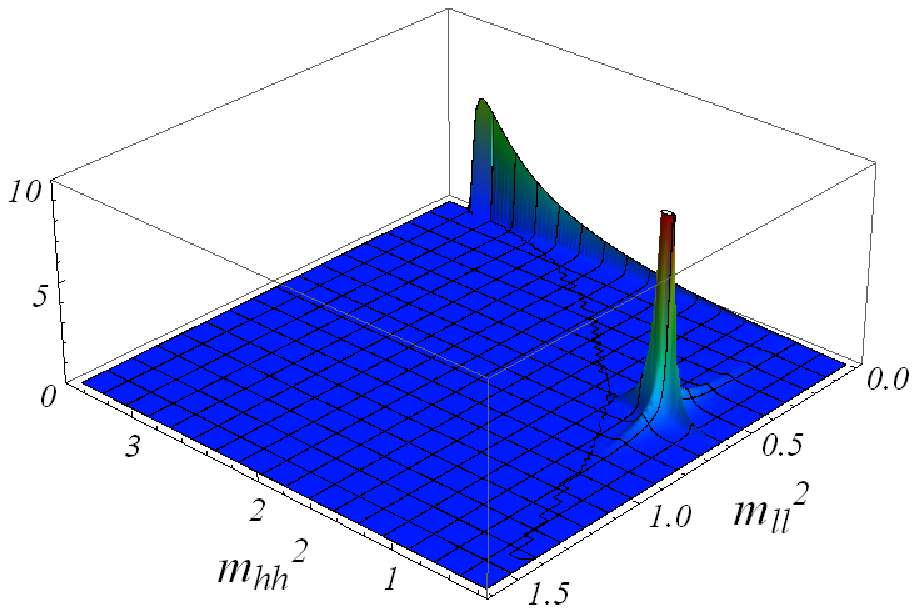}\\
\vspace{0.4cm}
\includegraphics[width=8.5cm]{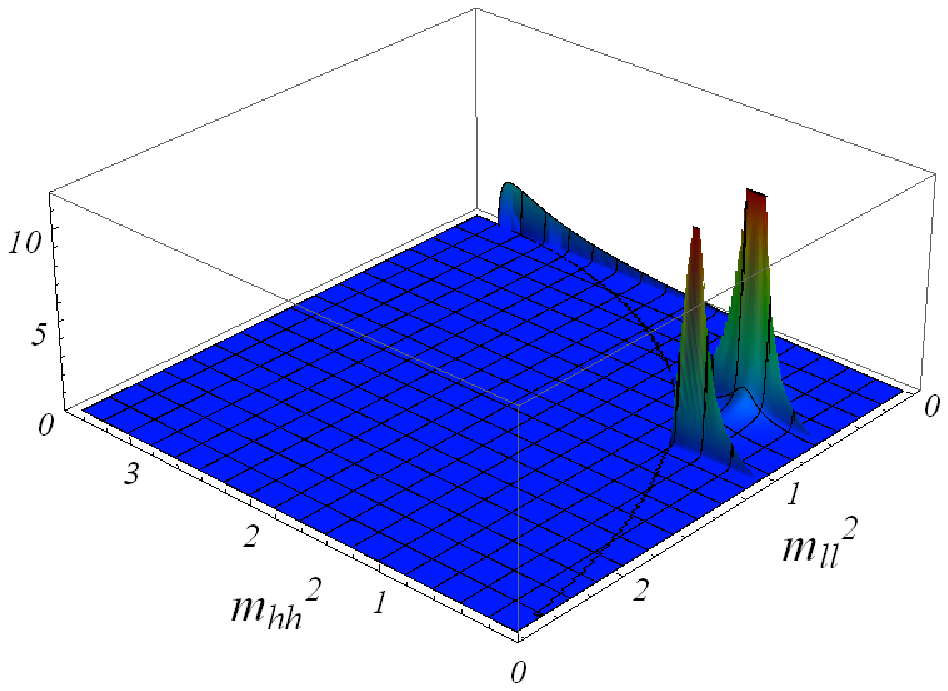}\\
\vspace{0.4cm}
\includegraphics[width=8.5cm]{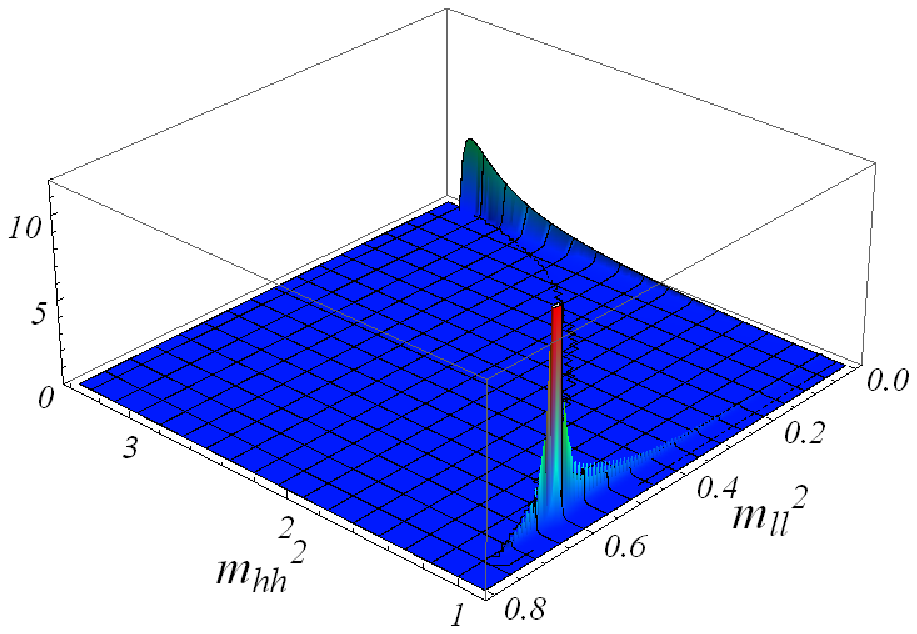}
\end{center}
\caption{\small{\it{Long-distance contributions to the differential decay width (in arbitrary units) in the $(m_{ll}^2,m_{hh}^2)$ plane for the different decay modes (from top to bottom: $K^{\mp}\pi^{\pm}$, $\pi^+\pi^-$ and $K^+K^-$, respectively). $m_{ll}$ and $m_{hh}$ are given in GeV. Above we show the $e^+e^-$ modes. For the dimuon case the only difference happens for the Bremsstrahlung, which is strongly suppressed due to lepton threshold effects.}}}\label{fig:3}
\end{figure}
\begin{align}\label{factorization}
\langle V\gamma^*|J^{\mu}_{ij} J_{\mu}^{{\bar{u}}c}| D^0\rangle&=\langle V|J^{\mu}_{ij}|0\rangle\langle\gamma^*|J_{\mu}^{{\bar{u}}c}|D^0\rangle\nonumber\\
&+\langle \gamma^*|J^{\mu}_{ij}|0\rangle\langle V|J_{\mu}^{{\bar{u}}c}|D^0\rangle\nonumber\\
&+\langle V\gamma^*|J^{\mu}_{ij}|0\rangle\langle0|J_{\mu}^{{\bar{u}}c}|D^0\rangle
\end{align}
where $ij={\bar{d}}d, {\bar{s}}s, {\bar{s}}d, {\bar{d}}s$. We will further assume that
\begin{itemize}
\item[(i)] The weak annihilation contribution (third line) is negligible compared to the spectator ones (first two lines) for all the processes to be considered (see, {\it{e.g.}},~\cite{Bauer:1986bm} for a detailed discussion).
\item[(ii)] The Zweig rule is at work, {\it{i.e.}}, flavor annihilation is suppressed and a possible enhancement due to final state interactions is excluded. 
\item[(iii)] The photon is created mainly through vector meson exchange and we will neglect a direct photon coupling. We will consider the exchange of the lowest-lying neutral states ($\rho,\omega,\phi$) to be dominant. 
\end{itemize}

The last point implies that
\begin{align}
\langle V\gamma^*|J^{\mu}_{ij} J_{\mu}^{{\bar{u}}c}| D^0\rangle&=\sum_{V^{\prime}}\langle \gamma^*|{\cal{H}}_{V\gamma}|V^{\prime}\rangle \frac{1}{P_{V^{\prime}}(q^2)}\nonumber\\
&\times\langle V^{\prime}V|J^{\mu}_{ij} J_{\mu}^{{\bar{u}}c}| D^0\rangle
\end{align}
and requires the electromagnetic couplings of vector mesons, which can be inferred from   
\begin{align}
{\cal{H}}_{V\gamma}&=-\frac{1}{4}\langle V_{\mu\nu}V^{\mu\nu}\rangle+\frac{f_Ve}{\sqrt{2}m_V}F_{\mu\nu}\langle Q V^{\mu\nu}\rangle
\end{align} 
where $f_V$ is defined by
\begin{align}
&\langle V(k,\epsilon)| J_{\mu} | 0\rangle= f_V m_V \epsilon^*_{\mu}(k)
\end{align}
For phenomenological purposes we will break the $SU(3)$ symmetry and take the experimental values for the decay couplings. The interaction term in the previous Hamiltonian will thus be replaced by
\begin{align}
{\cal{H}}_{V\gamma}&=-e\left(\frac{f_{\rho}}{m_{\rho}}\rho^{\mu}+\frac{f_{\omega}}{3m_{\omega}}\omega^{\mu}-\frac{\sqrt{2}f_{\phi}}{3m_{\phi}}\phi^{\mu}\right)\Box A_{\mu}
\end{align}

To complete the picture, we need the $\langle V(p)|J_{\mu}^{{\bar{u}}c}|D^0(P)\rangle$ matrix elements, which can be parameterized as ($p_+\equiv P+p$, $k\equiv P-p$)
\begin{align}
\langle V(p,\epsilon)|&J_{\mu}^{{\bar{u}}c}|D^0(P)\rangle=D_1(k^2)p_{+\mu}+D_2(k^2)k_{\mu}\nonumber\\
&+D_3(k^2)\epsilon_{\mu}^*+iD_4(k^2)\epsilon_{\mu\nu\lambda\rho}p_+^{\nu}k^{\lambda}\epsilon^{\rho*}
\end{align}
A determination of the different form factors $D_i(k^2)$ is given in Ref.~\cite{Wirbel:1985ji}. We refer to Appendix~\ref{sec:VII} for details. Matching Eqs.~(\ref{factorization}) and (\ref{formfactors}) the result for the $t_i(q^2,p^2)$ form factors is 
\begin{align}\label{formfactors1}
t_2^V(q^2,p^2)&=-\frac{2i\xi_V}{m_D+m_{\rho}}\nonumber\\
&\times\left[J^V(q^2){\hat{A}}_2(p^2)+\eta^V W(q^2){\hat{A}}_2(q^2)\right]\nonumber\\
t_3^V(q^2,p^2)&=i\xi_V(m_D+m_{\rho})\nonumber\\
&\times\left[J^V(q^2){\hat{A}}_1(p^2)+\eta^V W(q^2){\hat{A}}_1(q^2)\right]\nonumber\\
t_4^V(q^2,p^2)&=\frac{2\xi_V}{m_D+m_{\rho}}\nonumber\\
&\times\left[J^V(q^2){\hat{V}}(p^2)+\eta^V W(q^2){\hat{V}}(q^2)\right]
\end{align}
where $\xi_j=C_2^j\lambda_j\frac{eG_F}{\sqrt{2}}$, $\eta^V=0,1$ depending on the channel and
\begin{align}\label{JW}
J^V(q^2)&=q^2\left(\frac{f_{\rho}}{m_{\rho}P_{\rho}(q^2)}+\frac{f_{\omega}}{3m_{\omega}P_{\omega}(q^2)}\right)f_Vm_V\nonumber\\
W(q^2)&=q^2\left(\frac{f_{\rho}^2}{P_{\rho}(q^2)}+\frac{f_{\omega}^2}{3P_{\omega}(q^2)}-\frac{\sqrt{2}f_{\phi}^2}{3P_{\phi}(q^2)}\right)\nonumber\\
{\hat{A}}_2(k^2)&=\frac{h_{A2}m_{A2}^2}{m_{A2}^2-k^2}\nonumber\\
{\hat{A}}_1(k^2)&=\frac{h_{A1}m_{A1}^2}{m_{A1}^2-k^2}\nonumber\\ 
{\hat{V}}(k^2)&=\frac{h_{V1}m_{V1}^2}{m_{V1}^2-k^2}
\end{align}
As side remarks, we note that: (i) the first and second terms in Eq.~(\ref{formfactors1}) correspond to the first and second lines of Eq.~(\ref{factorization}), respectively. The absence of $\phi$-exchange in $J^V(q^2)$ is thus a consequence of the Zweig rule; and (ii) the global prefactor $(m_D+m_{\rho})$ in Eqs.~(\ref{formfactors1}) results from considering the $\rho(770)$ and $\omega(782)$ nearly degenerate.

Bringing all the pieces together and matching to Eq.~(\ref{param}) the form factors take the form:
\begin{align}
F_1^{(V)}&=i\frac{a_{21}^Vq\cdot p_1+a_{22}^Vq\cdot p_2+a_{31}^V}{P_V(p^2)}\nonumber\\
F_2^{(V)}&=i\frac{a_{21}^Vq\cdot p_1+a_{22}^Vq\cdot p_2+a_{32}^V}{P_V(p^2)}\nonumber\\
F_3^{(V)}&=\frac{a_4^V}{P_V(p^2)}
\end{align}
where
\begin{align}
a_{21}^V(q^2,p^2)&=-a_{22}^V(q^2,p^2)=-ib^Vt_2^V(q^2,p^2)\nonumber\\
a_{31}^V(q^2,p^2)&=-a_{32}^V(q^2,p^2)=-ib^Vt_3^V(q^2,p^2)\nonumber\\
a_4^V(q^2,p^2)&=-2b^Vt_4^V(q^2,p^2)
\end{align}


\subsection{Results and discussion}
In order to proceed to a numerical analysis for the different decay channels we need to estimate the parameters entering the different form factors. For the time being, and given the theoretical uncertainty in our results (which, while difficult to estimate, can easily amount to $30-50\%$), we will content ourselves with reference values for the different input parameters. 

In addition to the couplings in Eq.~(\ref{Vhh}) for the $V\to h_1^+h_2^-$ couplings, we will take~\cite{Ball:2006eu}
\begin{align}
f_{\rho}=216\,\, {\mathrm{MeV}},\qquad & f_{K^*}=220\,\, {\mathrm{MeV}},\nonumber\\
f_{\omega}=187\,\, {\mathrm{MeV}},\qquad & f_{\phi}=215\,\, {\mathrm{MeV}}
\end{align}
for the vector meson coupling constants. Regarding the ${\hat{A}}_2,{\hat{A}}_1,{\hat{V}}$ form factors in Eq.~(\ref{JW}), their residues have been determined experimentally for the $D^0\to K^*$ transition~\cite{PDG,Link:2004uk,Liu:2012bn}. In this work we will consider $r_V=h_{V1}/h_{A1}\sim 1.7$ and $r_2=h_{A2}/h_{A1}\sim 1$. Additionally, we will assume that $h_{A1}=0.55$~\cite{ElHassanElAaoud:1999nx}, which leads to $h_{V1}=0.94$ and $h_{Ai}=0.55$. We will consider the previous numbers to be flavor-blind. Regarding the poles, we will adopt the usual values, namely $m_{V1}=2110$ MeV and $m_{A1}=m_{A2}=2530$ MeV~\cite{Wirbel:1985ji}. We will also assume that the Wilson coefficient $C_2\simeq -0.55$ for all the channels~\cite{Bauer:1986bm}. The remaining parameters (CKM matrix elements, decay widths and vector meson masses) are taken from~\cite{PDG}. 

The main results are collected in Table~\ref{tab1} and Fig.~\ref{fig:3}. In Table~\ref{tab1} we have listed the contributions of Bremsstrahlung, electric and magnetic emission to the branching ratios for the different channels (the electric-Bremsstrahlung interference is only at the $1-2\%$ level and has been omitted). The amplitudes turn out to be largely dominated by the electric pieces (Bremsstrahlung and electric emission for the $e^+e^-$ decays, or only electric emission for the $\mu^+\mu^-$ decays) with weights that depend on the channel under study. Fig.~\ref{fig:3} shows the differential decay widths for the different channels in the $(m_{ll}^2,m_{hh}^2)$ plane. Analytical expressions thereof can be found in Appendix~\ref{sec:VIII}. Notice that, in agreement with Low's theorem, the low-$q^2$ region is dominated by the Bremsstrahlung. The resonance contributions (or any other contributions) should have a smooth low-$q^2$ behavior. Therefore, we do not quite understand the low-$q^2$ enhancement found in Ref.~\cite{Fajfer:1998rz} for $D^0\to V{\ell}^+{\ell}^-$, but it definitely cannot be ascribed to factorization, as suggested in~\cite{Burdman:2001tf}.\footnote{Apparently, the low-$q^2$ enhancement was corrected in~\cite{Fajfer:2005ke} (S. Fajfer, private communication).}

In the following we will discuss each separate channel in turn, focussing on their specific features.


\subsubsection{$D^0\to K^-\pi^+ l^+l^-$}
The Bremsstrahlung contribution is listed in Appendix~\ref{sec:VIII}. The dominant resonance contribution to this Cabibbo-allowed decay comes from $D^0\to {\overline{K^*}}V^{\prime}$, with subsequent decays ${\overline{K^*}}\to K^-\pi^+$ and $V^{\prime}\to \gamma^*\to l^+l^-$. One can easily get convinced that strangeness conservation forbids the second line in the factorization formula of Eq.~(\ref{factorization}). The form factors get therefore reduced to 
\begin{align}\label{formsimple}
t_2^{K^*}(q^2,p^2)&=-\frac{2i\xi_{K^*}}{m_D+m_{\rho}}J^{K^*}(q^2){\hat{A}}_2(p^2)\nonumber\\
t_3^{K^*}(q^2,p^2)&=i\xi_{K^*}(m_D+m_{\rho})J^{K^*}(q^2){\hat{A}}_1(p^2)\nonumber\\
t_4^{K^*}(q^2,p^2)&=\frac{2\xi_{K^*}}{m_D+m_{\rho}}J^{K^*}(q^2){\hat{V}}(p^2)
\end{align}
The combination of long-distance contributions is shown in the top panel of Fig.~\ref{fig:3}, where one can see the hadronic emission sitting (mostly) at $(m_{\rho}^2,m_{K^*}^2)$. The $\omega(782)$ contributions is also present but too small to be noticed by the naked eye. As expected, the Bremsstrahlung and resonance regions lie rather far apart, which explains why their interference is negligible. For the $e^+e^-$ case, the total branching ratio adds up to $1.6\cdot 10^{-5}$, with a slight dominance of the Bremsstrahlung, while for $\mu^+\mu^-$ we find $6.7\cdot 10^{-6}$, coming largely from the electric emission.

The same qualitative features apply for the doubly Cabibbo-suppressed modes. In this case, branching ratios add to $5.5\cdot 10^{-8}$ for electrons and $1.8\cdot 10^{-8}$ for muons.   

  
\subsubsection{$D^0\to\pi^+\pi^- l^+l^-$}

The direct emission contribution for the $D^0\to\pi^+\pi^-\gamma^*$ decay comes mainly from the near-resonant decays $D^0\to \rho^0\gamma^*$ and $D^0\to\omega\gamma^*$. As we argued in the previous subsection, for phenomenological purposes it is a good approximation to neglect the $\omega$ contribution due to its tiny branching ratio into two pions.

Fig.~\ref{fig:3} shows both Bremsstrahlung and direct emission contributions. The resonance region now shows distinct peaks at $(m_{\rho}^2, m_{\rho}^2)$ and $(m_{\phi}^2,m_{\rho}^2)$ of comparable size. Contrary to the Cabibbo-allowed mode, in this case both spectator pieces contribute and Eq.~(\ref{formfactors1}) assumes its more general form for the $t_i^{\rho}$.

The branching ratios for the $e^+e^-$ and $\mu^+\mu^-$ are respectively $2.0\cdot 10^{-6}$ and $1.5\cdot 10^{-6}$. We note that our results for the direct emission are in very good agreement with Ref.~\cite{Burdman:2001tf}, where the 3-body decays $D^0\to \rho l^+l^-$ were considered. Our results are especially interesting in view of the prospects of LHCb to be able to detect signals at $10^{-6}$ in Cabibbo-suppressed $D^0\to h_1^+h_2^- \mu^+\mu^-$ decays~\cite{Bediaga:2012py}. 


\subsubsection{$D^0\to K^+K^- l^+l^-$}

The direct emission contribution comes in this case mainly from a near-resonant $D^0\to \phi\gamma^*$. Similarly to the Cabibbo-allowed case, only the first line survives in factorization and therefore the form factors reduce to Eq.~(\ref{formsimple}), with the obvious substitution $K^*\to \phi$. 

The resonance region is peaked at $(m_{\rho}^2,m_{\phi}^2)$, such that Bremsstrahlung and direct emission populate opposite sides of the phase space in the Dalitz plot. The total branching ratios for the $e^+e^-$ and $\mu^+\mu^-$ are respectively $6.5\cdot 10^{-7}$ and $1.1\cdot 10^{-7}$, with a slight dominance of the Bremsstrahlung for the electron case.


\section{Short-distance effective Hamiltonian}\label{sec:IV}
In the previous Sections we have studied the (dominant) long-distance contributions to the $D^0\to h_1^+h_2^-{\ell}^+{\ell}^-$ decays. In this Section we turn our attention to short distances. Our main aim is to single out an observable without long-distance background. In charm decays, such an object is automatically a new physics probe, since the standard model contribution is not only loop-suppressed, but additionally shrinked by the GIM mechanism. In the following, we will first assess the standard model prediction for $D^0\to h_1^+h_2^-{\ell}^+{\ell}^-$ and then discuss new physics scenarios with enhanced semileptonic operators and their connection with $\Delta a_{CP}$.


\subsection{Standard model prediction} 

The $c\to u l^+l^-$ transition inside the standard model is mediated by electromagnetic and Z penguin diagrams and by W box diagrams at one-loop order. At energies right below the charm threshold it is described by the following effective lagrangian
\begin{align}
{\cal{H}}_{eff}^{(c\to ul^+l^-)}&={\cal{H}}_{eff}^{(c\to u \gamma)}-\frac{G_F}{\sqrt{2}}\sum_{i}\lambda_i\left[C_9^{(i)} Q_9+ C_{10}^{(i)} Q_{10}\right]
\end{align}
where $\lambda_i=V_{ci}^*V_{ui}$, $i=d,s$ run through the down-type quarks and
\begin{align}
Q_9&=({\bar{u}}\gamma_{\mu}P_Lc)({\bar{{\ell}}}\gamma^{\mu}{\ell})\nonumber\\
Q_{10}&=({\bar{u}}\gamma_{\mu}P_Lc)({\bar{{\ell}}}\gamma^{\mu}\gamma_5{\ell})
\end{align}
The full list of operators in ${\cal{H}}_{eff}^{(c\to u \gamma)}$ can be found in~\cite{Burdman:1995te}. The leading contribution in $c\to u\gamma$ comes from the electromagnetic penguin operator $Q_{7L}$
\begin{align}
Q_{7L}&=i\frac{e^2}{4\pi^2}m_c\frac{q^{\nu}}{q^2}({\bar{u}}\sigma_{\mu\nu}P_Rc)({\bar{{\ell}}}\gamma^{\mu}{\ell})
\end{align}
The remaining operators are ${\cal{O}}(\alpha_s)$-suppressed~\cite{Buchalla:1995vs} and will be neglected. We have also omitted $Q_{7R}$ and $Q_{9,10}^{\prime}$, which are weighted by $m_u$ and can be safely neglected.

The RG running of the previous operators was first computed in~\cite{Inami:1980fz} to the leading logarithm approximation. QCD corrections to $c\to u\gamma$ were shown to lead to a significant enhancement of $C_{7L}$~\cite{Burdman:1995te}, where the strongest effect comes from the two-loop contribution: mixing with mainly $Q_2$ softens the GIM suppression from power-like to logarithmic~\cite{Greub:1996wn}. This phenomenon was first pointed out in~\cite{Shifman:1976de} and later applied to B decays~\cite{Bertolini:1986th} and kaon decays~\cite{Dib:1990gr,Riazuddin:1993pn}. Despite this enhancement, when it comes to $c\to ul^+l^-$, $Q_{7L}$ is overshadowed by $Q_9$ at least by one order of magnitude, since its mixing with $Q_2$ happens already at tree level~\cite{Buchalla:1995vs}. Although one-loop QCD corrections are significant~\cite{Fajfer:2002gp} and effectively reduce the value of $C_9$, it still constitutes the dominant contribution to the short-distance standard model estimate of $c\to ul^+l^-$~\cite{Paul:2011ar}. In contrast, $Q_{10}$ does not mix with $Q_2$ and is invariant under the renormalization flow, which makes its value extremely suppressed in the standard model:
\begin{align}
C_{10}(m_c)&=C_{10}(m_W)\sim \frac{m_s^2}{m_W^2}
\end{align}
For inclusive $c \to ul^+l^-$ transitions, the branching ratio has recently been estimated at $3.7 \cdot 10^{-9}$ for electron-positron dilepton pair~\cite{Paul:2011ar}. For the dimuon case, it is reasonable to expect a factor 5 suppression. Taking into account our results in table~\ref{tab1}, this entails that short-distance effects have a negligible impact on the decay width, as expected. Even if one is extremely conservative and makes the comparison only with the Bremsstrahlung, short distances might be competitive only for the dimuon Cabibbo-suppressed modes. However, disentangling short from long distances in those cases is extremely challenging: since the bulk of the short distances comes from $C_9$, the short-distance piece has the same distribution in the Dalitz plot as the long-distance one, {\it{i.e.}}, the short-distance contribution will simply pile up on top of the long-distance contribution.


\subsection{New physics scenarios}\label{newphysics}
In this section we will discuss two different new physics scenarios, namely SUSY in the single-insertion approximation and generic Z-enhanced models, that can generate $\Delta a_{CP}$ at the experimentally observed level through enhanced magnetic penguins. Our main goal will be to determine the typical sizes they induce for semileptonic operators. 

In SUSY models within the single-insertion approximation, the s-quark gluino loop can easily account for $\Delta a_{CP}$ if~\cite{Isidori:2012yx} 
\begin{align}\label{C7}
|\mathrm{Im}C_{7,8}(m_c)|\sim 4\cdot 10^{-3}
\end{align}
Quite generally one can show that, for the hadronic part of the $D^0\to 2h2l$ decay,  
\begin{align}
q^{\nu}\langle \pi^+\pi^-|{\bar{u}}\sigma_{\mu\nu}P_Rc|D^0\rangle=i\eta_T q^2\langle \pi^+\pi^-|{\bar{u}}\gamma_{\mu}P_Lc|D^0\rangle
\end{align}
where $\eta_T$ parametrizes the relative strength of the vector and tensor matrix elements. Its magnitude can be estimated with different hadronic models, but one naturally expects that $\eta_T\sim {\cal{O}}(1)$. Using the definitions of $Q_{7L}$ and $Q_9$ given in the previous section, one concludes that
\begin{align}
\langle 2h2l|Q_{7L}|D^0\rangle=-\frac{\alpha}{\pi}m_c\eta_T\langle 2h2l|Q_{9}|D^0\rangle
\end{align}
While the previous relation holds at the operator level, in practice one expects that $C_9$ will be suppressed for chirality reasons. 

Larger values for both real and imaginary parts of $C_9$ and $C_{10}$ can be generated with other mechanisms, for instance with double-insertion diagrams correcting the $Z$ vertex ({\it{e.g.}} the s-quark gluino loop). In order to be more general, one can parametrize these new physics effects correcting Z vertices in an effective field theory language. These Z-enhanced scenarios are governed by the Lagrangian:  
\begin{align}
{\cal{L}}_{NP}=-\frac{g}{2c_W}{\bar{q}}_i\gamma_{\mu}\left[g_L^{ij}P_L+g_R^{ij}P_R\right]q_jZ^{\mu}
\end{align} 
Semileptonic operators receive contributions from tree-level Z-exchange diagrams, with the result:
\begin{align}
C_9^{NP}&=-\frac{g_{L}^{uc}}{\lambda_b}(1-4s_W^2)\nonumber\\
C_{10}^{NP}&=\frac{g_{L}^{uc}}{\lambda_b}
\end{align}
$D^0-{\bar{D^0}}$ mixing puts stringent bounds on the up-type couplings, $|g_{L}^{uc}|< 2\cdot 10^{-4}$~\cite{Giudice:2012qq}. Assuming $g_{L}^{uc}$ to be real, this translates into $C_9^{NP}<0.1$ and $C_{10}^{NP}<1.25$. In~\cite{Giudice:2012qq} it was observed that with $g^{uc}$ couplings alone magnetic penguins cannot be enhanced to fit $\Delta a_{CP}$. However, one can account for $\Delta a_{CP}$ if one-loop diagrams with top exchange are sizeable. This is feasible because the top sector is only loosely constrained ($|g_{L}^{ut}|<2\cdot 10^{-2}$). For the semileptonic operators, such one-loop contributions will typically be of the form
\begin{align}
\frac{ C^{1-loop}_{9,10}}{C_{9,10}}\sim \frac{g_L^{ut}(g_L^{ct})^*}{g_L^{uc}}\sim {\cal{O}}(1)
\end{align}  
As we will discuss in the following section, new physics in 4-body $D^0$ decays is not restricted to CP-violating observables. In particular, one can define clean observables for new physics sensitive to the real parts of $C_{9,10}$, hence not constrained by $\Delta a_{CP}$. In the following, we will assume that the CP-violating phases entering $\Delta a_{CP}$ come mainly from the right-handed Z couplings while the left-handed couplings are mainly real and generate $C_{9,10}\sim {\cal{O}}(1)$. Incidentally, we note that this complies with the values adopted in Little Higgs Model scenarios~\cite{Fajfer:2005ke,Paul:2011ar}. In particular, Ref.~\cite{Fajfer:2005ke} used $C_{9,10}=4$. In the following we will adopt a more conservative $C_{9,10}=1$.

\begin{table*}[t]
\begin{center}
\begin{tabular}{|ccc|}
\hline
Mode &\,\,\,\, $(e^+e^-)_{\phi}$ &\,\,\,\, $(\mu^+\mu^-)_{\phi}$\\
\hline
$K^-\pi^+$ &\,\,\,\,$5.4\cdot 10^{-7}\,(\sim 3\%)$ &\,\,\,\, $4.8\cdot 10^{-7}\,(\sim 7\%)$\\
$\pi^+\pi^-$ &\,\,\,\,$1.3\cdot 10^{-7}\,(\sim 6\%)$ &\,\,\,\, $1.1\cdot 10^{-7}\,(\sim 8\%)$\\
$K^+K^-$ &\,\,\,\,$7.9\cdot 10^{-9}\,(\sim 1\%)$ &\,\,\,\, $6.9\cdot 10^{-9}\,(\sim 6\%)$\\
$K^+\pi^-$ &\,\,\,\,$1.5\cdot 10^{-9}\,(\sim 3\%)$ &\,\,\,\, $1.3\cdot 10^{-9}\,(\sim 7\%)$\\
\hline
\end{tabular}
\end{center}
\caption{\small{\it{Branching ratios for the interference between the electric and magnetic terms assuming $\delta_W\simeq \pi/4$ for the weak phase. In parenthesis we show the value for the T-odd asymmetry $A_{\phi}$.}}}\label{tab2}
\end{table*}


\section{Angular asymmetries}\label{sec:III}
The most general angular distribution for a 4-body decay can be parametrized, using the Cabibbo-Maksymowicz set of variables, as in Eq.~(\ref{angular}), which we repeat here for convenience:
\begin{align}\label{angular2}
\frac{d^5\Gamma}{dxdy}&={\cal{A}}_1(x)+{\cal{A}}_2(x)s_{\ell}^2+{\cal{A}}_3(x)s_{\ell}^2c_{\phi}^2+{\cal{A}}_4(x) s_{2{\ell}}c_{\phi}\nonumber\\
&+{\cal{A}}_5(x) s_{\ell}c_{\phi}+{\cal{A}}_6(x)c_{\ell}+{\cal{A}}_7(x) s_{\ell}s_{\phi}\nonumber\\
&+{\cal{A}}_8(x)s_{2\ell}s_{\phi}+{\cal{A}}_9(x)s_{\ell}^2s_{2\phi}
\end{align}
The pieces contributing to the decay width were studied in Section~\ref{sec:II} and correspond to the first line above. The remaining angular structures can be generated by interference effects, either at the hadronic or at the leptonic vertex, and can be probed with different angular asymmetries. It is important to emphasize that, due to the rich angular structure of Eq.~(\ref{angular2}), angular asymmetries are not restricted to charge asymmetries. Moreover, there are asymmetries that, due to tiny standard model backgrounds, turn out to be clean probes of new physics. In this Section we will concentrate on two such asymmetries. Far from being exhaustive, we just want to provide some exploratory indications of what are their expected signals in the Dalitz plot and parametrize its size in terms of a few short-distance parameters. 

\begin{figure*}[h!]
\begin{center}
\includegraphics[width=7.5cm]{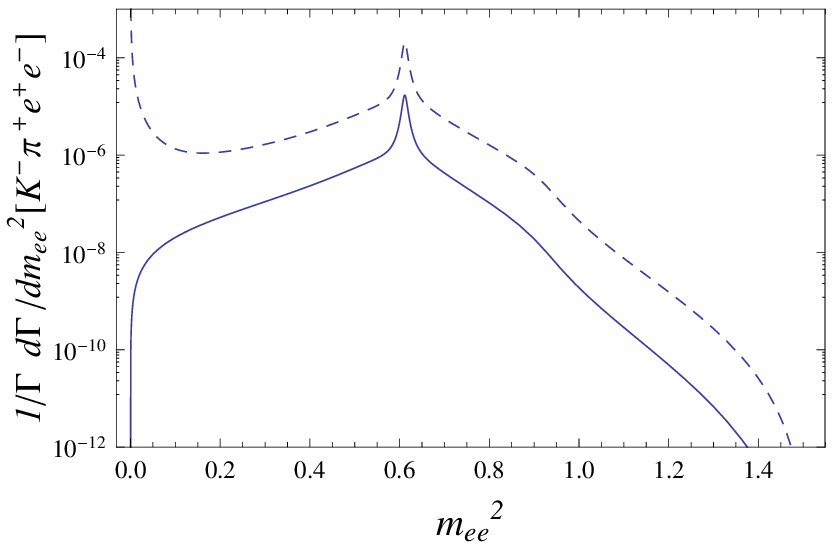}\hspace{0.3cm}\includegraphics[width=7.5cm]{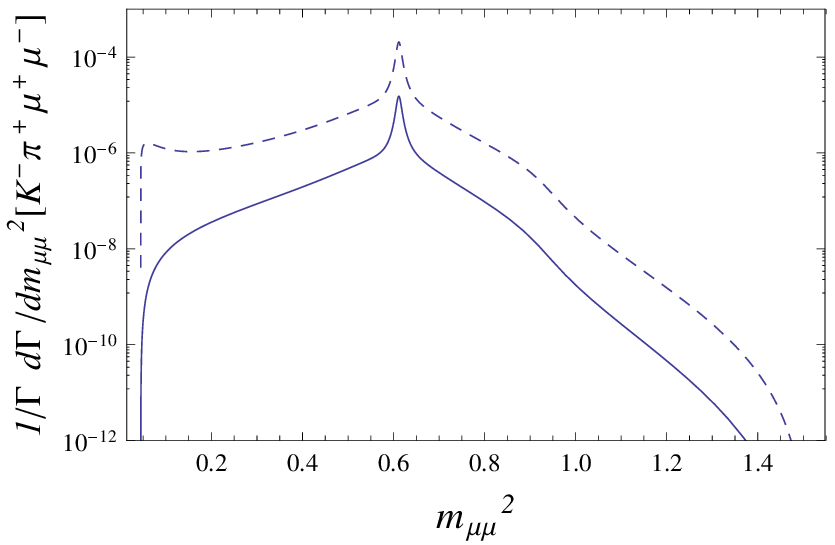}\\
\vspace{0.4cm}
\includegraphics[width=7.5cm]{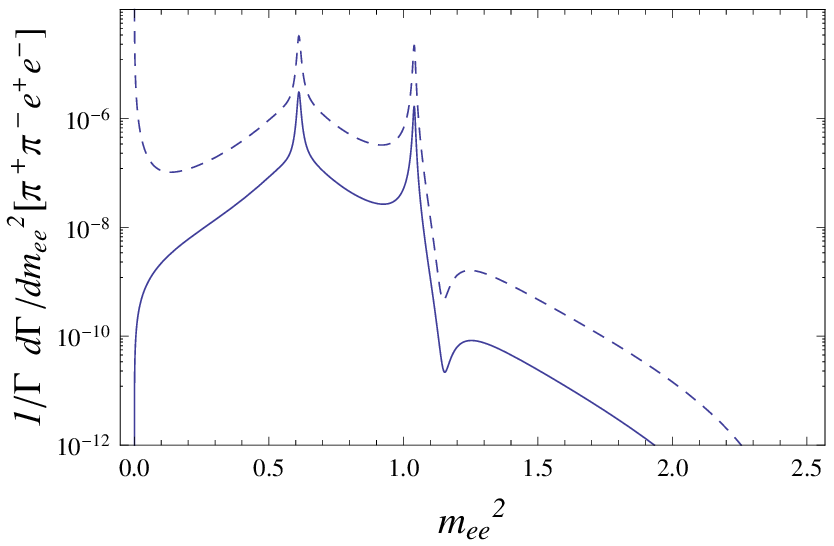}\hspace{0.3cm}\includegraphics[width=7.5cm]{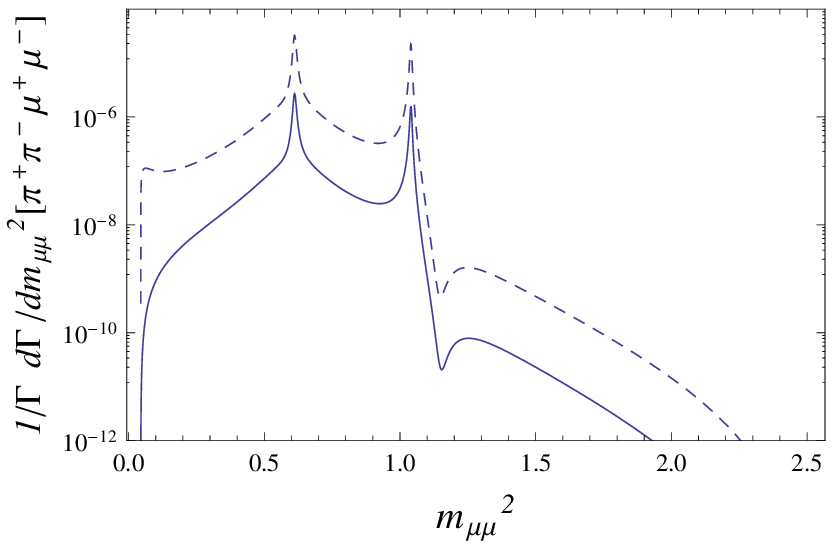}\\
\vspace{0.4cm}
\includegraphics[width=7.5cm]{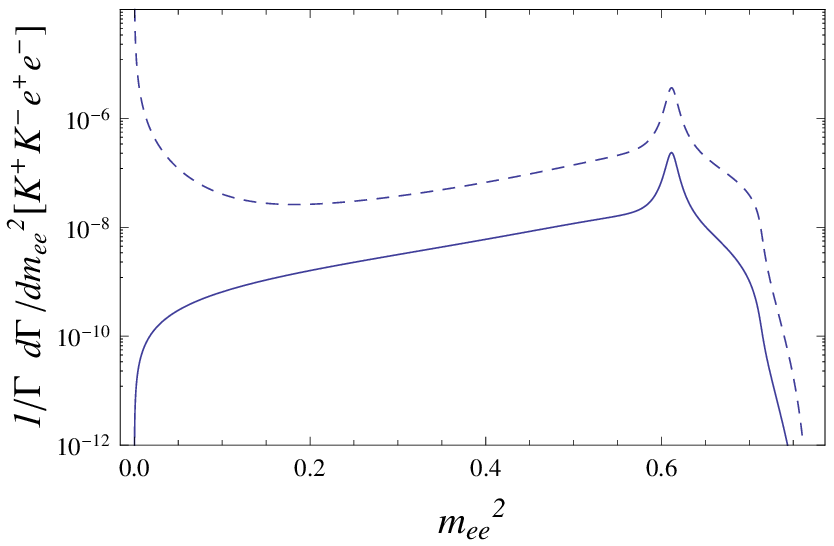}\hspace{0.3cm}\includegraphics[width=7.5cm]{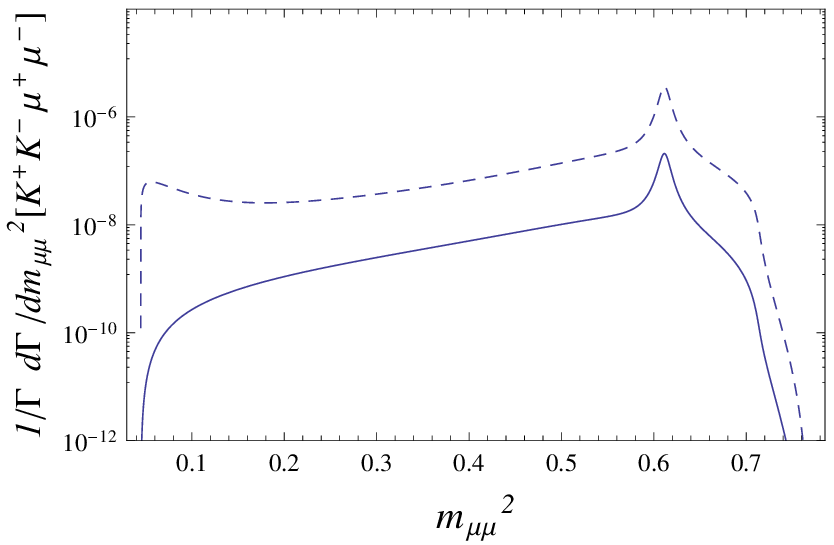}\\
\vspace{0.4cm}
\includegraphics[width=7.5cm]{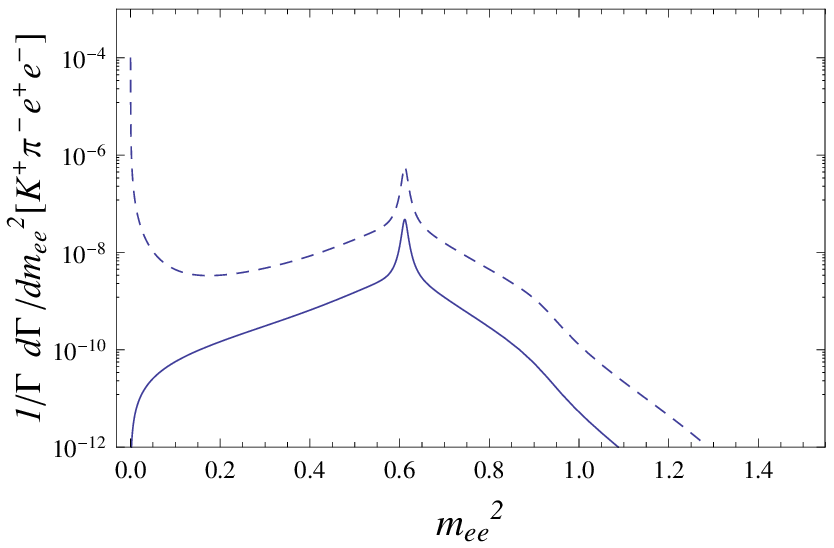}\hspace{0.3cm}\includegraphics[width=7.5cm]{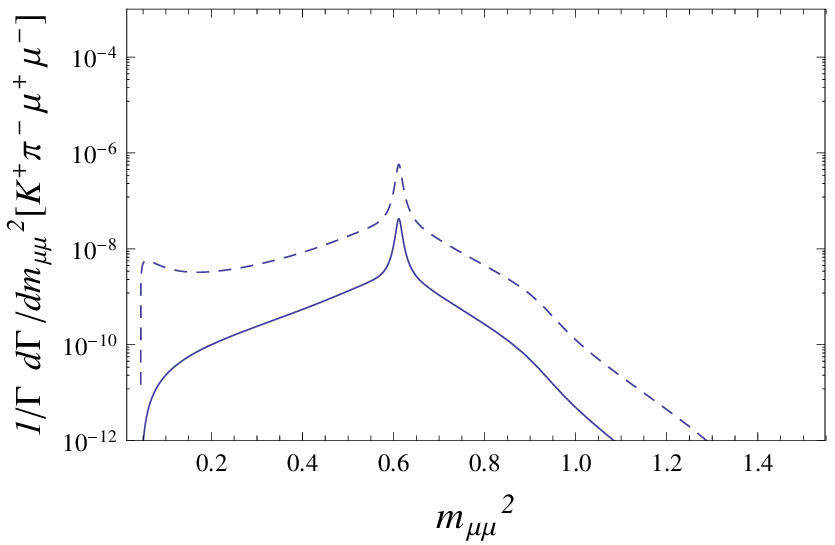}\\
\end{center}
\caption{\small{\it{The angular integrated (with the prescription of Eq.~(\ref{angularasymm})) differential decay width as a function of $m_{ll}^2$ (in GeV$^2$) for the different decay modes (solid lines). The vertical axis displays $\displaystyle\frac{d\Gamma^*}{dm_{ll}^2}\equiv \int_0^{2\pi} \frac{d\Gamma}{dm_{ll}^2d\phi}d\phi^*$, which corresponds to Eq.~(\ref{C10}) integrated over $p^2$. For concreteness, $\delta_{W}=\pi/4$ is chosen as a reference value. For convenience the result is normalized to the total decay width, such that the area under the curve is $A_{\phi}$ (see Eq.~(\ref{Plong})). The dashed lines correspond to the (angular symmetric) differential decay width, which is included for comparison. From top to bottom: $K^-\pi^+$, $\pi^+\pi^-$, $K^+K^-$ and $K^+\pi^-$ modes. Left and right panels collect, respectively, the $e^+e^-$ and $\mu^+\mu^-$ modes.}}}\label{fig:4}
\end{figure*} 

\subsection{T-odd asymmetry} 
It was long noted, in the context of $K_L\to \pi^+\pi^-e^+e^-$, that the interference between the magnetic and Bremsstrahlung contributions can lead to large CP violation~\cite{Sehgal:1992wm}. This interference is genuine of 4-body decays and contributes to ${\cal{A}}_{8,9}$ in Eq.~(\ref{angular2}). Therefore, it can be singled out by an angular asymmetry in the diplane angle $\phi$:   
\begin{align}\label{Plong}
A_{\phi}&=\langle {\mathrm{sgn}}(s_{\phi}c_{\phi})\rangle=\frac{1}{\Gamma}\int_0^{2\pi}\frac{d\Gamma}{d\phi}d\phi^*
\end{align}
where we have defined the piece-wise angular integration 
\begin{equation}\label{angularasymm}
\int_0^{2\pi}d\phi^*\equiv \left[\int_0^{\pi/2}-\int_{\pi/2}^{\pi}+\int_{\pi}^{3\pi/2}-\int_{3\pi/2}^{2\pi}\right]d\phi
\end{equation}
This asymmetry actually collects not only the Bremsstrahlung {\it{vs.}} magnetic interference, but also the electric {\it{vs.}} magnetic one. In kaon physics the former dominates and, since the Bremsstrahlung and magnetic contributions have different strong phases, $A_{\phi}$ becomes a probe of long-distance physics (see, {\it{e.g.}},~\cite{Cappiello:2011qc}).

In $D$ decays, in contrast, the electric {\it{vs}} magnetic interference is expected to be dominant, because both pieces have the same structure in the Dalitz plot. In comparison, the Bremsstrahlung {\it{vs}} magnetic interference here is severely suppressed, even allowing for large strong phases. Interestingly, the electric {\it{vs}} magnetic interference can only be nonzero in the presence of weak phases. In Section~\ref{sec:II} we evaluated the hadronic matrix elements for the dominant $Q_2$ four-quark operator. Weak phases emerge when other four-quark operators are also considered, such as $Q_1$ or the QCD penguins, so that their presence is naturally expected. In the standard model, however, they are extremely suppressed and can only become sizeable in the presence of new physics. Therefore, as opposed to kaon physics, in $D$ physics $A_{\phi}$ is a probe of new physics, with signals mostly concentrated on the resonance region of the Dalitz plot. 
  
In Fig.~\ref{fig:4} we show the differential electric-magnetic interference as a function of $m_{ll}^2$ for the different decay modes. For comparison, we have included the long-distance background (dashed line). For the sake of illustration we have picked $\delta_W\sim \pi/4$ as a reference value for the weak phases. As expected, the contributions are sizeable close to the exchanged-resonance peaks, which vary depending on the final dihadron state. The interference between magnetic and Bremsstrahlung (not shown in the plot) amounts to roughly $1\%$ of the integrated asymmetry and can be thus safely neglected. In Table~\ref{tab2} we have listed the resulting values for $A_{\phi}$ for $\delta_W\sim \pi/4$. Notice that in that case the asymmetry hovers in the ${\cal{O}}(1-10\%)$ window, depending on the channel. Quite generically, dimuon modes give bigger signals than $e^+e^-$ modes, which should be taken as an additional motivation to study the dimuon decays by LHCb, specifically $D^0\to \pi^+\pi^-\mu^+\mu^-$. 


\subsection{Forward-Backward asymmetry}
In the previous subsection P violation was induced in the hadronic vertex. It is also interesting to consider P violation in the leptonic vertex.   

Consider the matrix element stemming from the semileptonic operators $Q_9$ and $Q_{10}$, namely
\begin{align}\label{amplitude2}
{\cal M}_{SD}^{(9)}\equiv\xi_9 L^{\mu}(k_+,k_-) {\cal{H}}_{\mu}(p_1,p_2,q)\nonumber\\
{\cal M}_{SD}^{(10)}\equiv\xi_{10}L^{\mu 5}(k_+,k_-) {\cal{H}}_{\mu}(p_1,p_2,q)
\end{align} 
where
\begin{align}
L^{\mu}(k_+,k_-)=\bar{u}(k_-)\gamma^\mu v(k_+)\nonumber\\
L^{\mu 5}(k_+,k_-)=\bar{u}(k_-)\gamma^\mu\gamma^5 v(k_+)
\end{align}
The short-distance hadronic tensor is defined as
\begin{align}\label{paramSD}
{\cal{H}}_{\mu}(p_1,p_2,q)&\equiv\langle h_1h_2|J_{\mu}^{{\bar{u}}c}|D^0\rangle
\end{align}
and
\begin{align}
\xi_{(9,10)}=\frac{G_F}{\sqrt{2}}\lambda_b C_{(9,10)}
\end{align}
Above, the CKM unitarity relation $\lambda_d+\lambda_s+\lambda_b=0$ has been used. The short-distance hadronic vector
\begin{figure*}[t]
\begin{center}
\includegraphics[width=7.5cm]{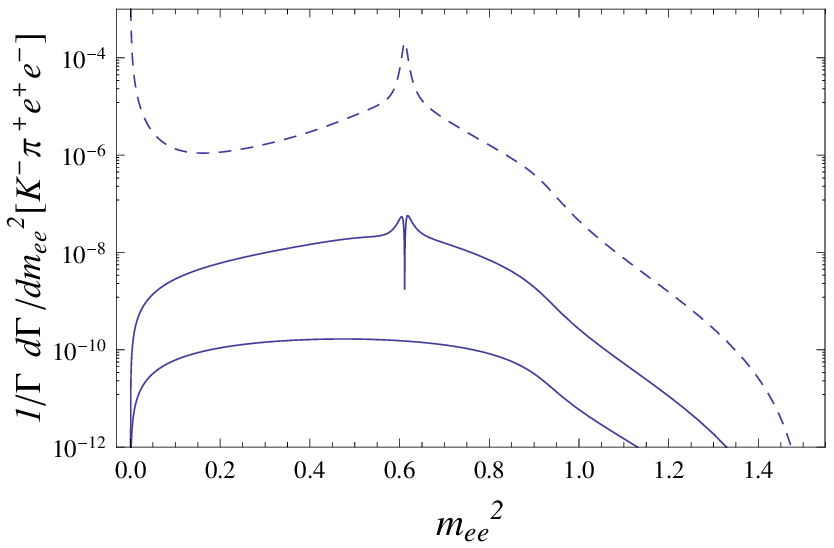}\hspace{0.3cm}\includegraphics[width=7.5cm]{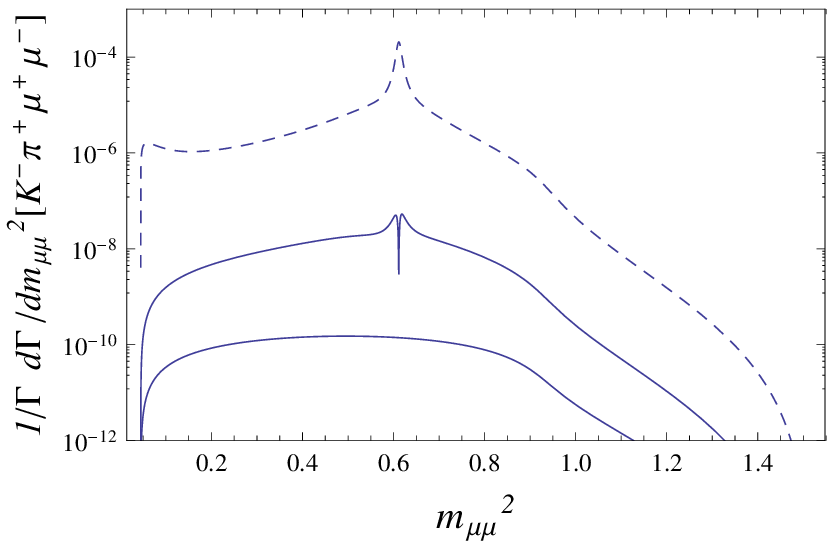}\\
\vspace{0.4cm}
\includegraphics[width=7.5cm]{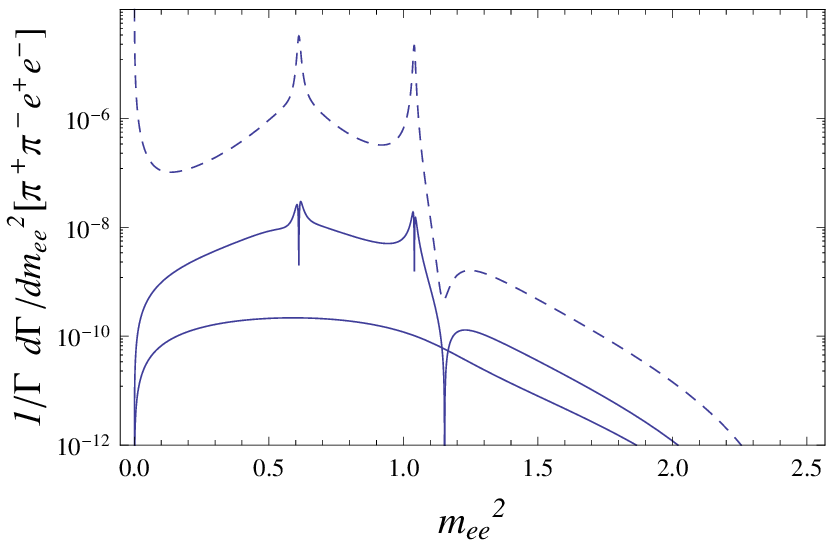}\hspace{0.3cm}\includegraphics[width=7.5cm]{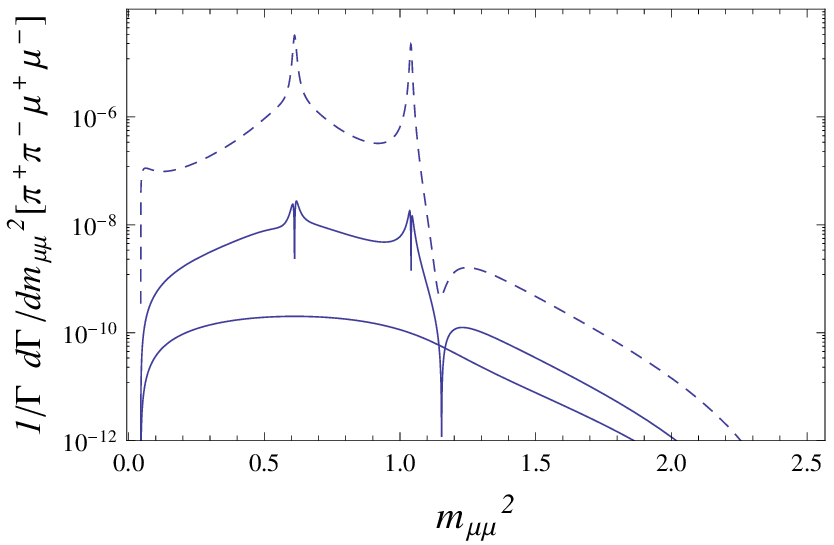}\\
\vspace{0.4cm}
\includegraphics[width=7.5cm]{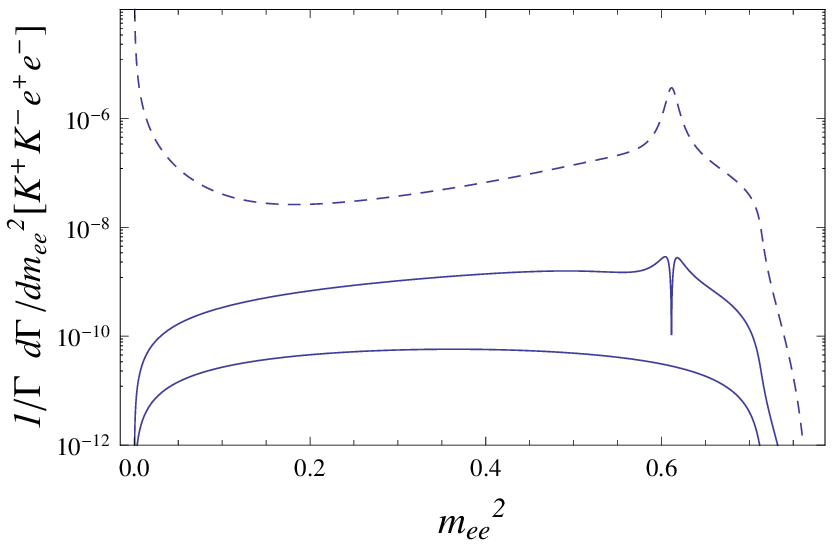}\hspace{0.3cm}\includegraphics[width=7.5cm]{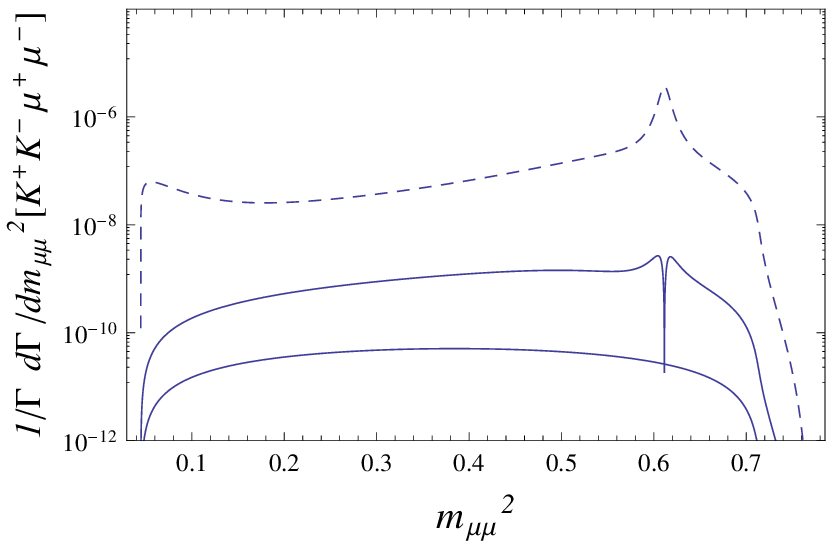}\\
\vspace{0.4cm}
\includegraphics[width=7.5cm]{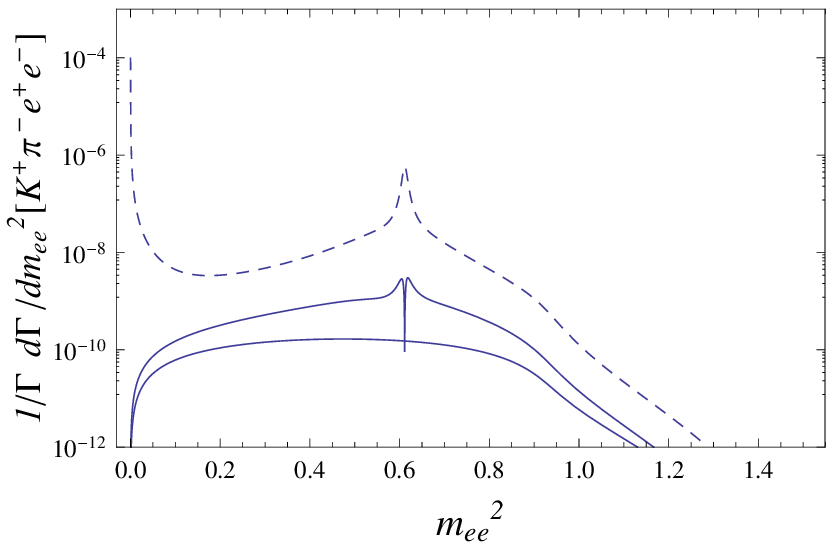}\hspace{0.3cm}\includegraphics[width=7.5cm]{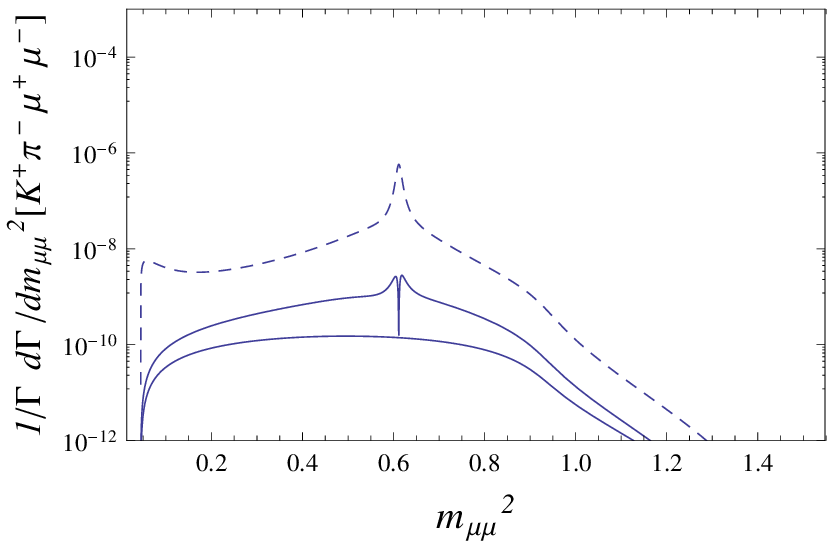}\\
\end{center}
\caption{\small{\it{
The angular integrated (with the prescription of Eq.~(\ref{forback})) differential decay width as a function of $m_{ll}^2$ (in GeV$^2$) for the different decay modes. The vertical axis displays $\displaystyle\frac{d\Gamma_{FB}}{dm_{ll}^2}\equiv \bigg[\int_0^{1}-\int_{-1}^{0}\bigg] \frac{d\Gamma}{dm_{ll}^2dy}dy$. The spiky and smooth solid lines correspond, respectively, to the absolute value of Eq.~(\ref{C13}) and Eq.~(\ref{C14}) integrated over $p^2$. For concreteness, we have chosen $C_{9}=C_{10}=1$ as reference values. For convenience the result is normalized to the total decay width, such that the area under the curve is $A_{FB}$ (see Eq.~(\ref{forback})). The dashed lines correspond to the (angular symmetric) differential decay width, which is included for comparison. From top to bottom: $K^-\pi^+$, $\pi^+\pi^-$, $K^+K^-$ and $K^+\pi^-$ modes. Left and right panels collect, respectively, the $e^+e^-$ and $\mu^+\mu^-$ modes.}}}\label{fig:5}
\end{figure*}
\begin{table*}[t]
\begin{center}
\begin{tabular}{|ccc|cc|}
\hline
Mode &\,\,\,\, $(e^+e^-)_{SD-LD}$ &\,\,\,\, $(\mu^+\mu^-)_{SD-LD}$&\,\,\,\, $(e^+e^-)_{SD-SD}$ &\,\,\,\, $(\mu^+\mu^-)_{SD-SD}$\\
\hline
$K^-\pi^+$ &\,\,\,\,$1.1\cdot 10^{-8}\,(\sim 0.07\%)$ &\,\,\,\, $1.0\cdot 10^{-8}\,(\sim 0.06\%)$ &\,\,\,\, $1.0\cdot 10^{-10}\,(\sim 6\cdot 10^{-4}\%)$ &\,\,\,\, $7.1\cdot 10^{-11}\,(\sim 4\cdot 10^{-4}\%)$\\
$\pi^+\pi^-$ &\,\,\,\,$7.1\cdot 10^{-9}\,(\sim 0.4\%)$ &\,\,\,\, $6.5\cdot 10^{-9}\,(\sim 0.5\%)$ &\,\,\,\, $1.3\cdot 10^{-10}\,(\sim 7\cdot 10^{-3}\%)$ &\,\,\,\, $1.0\cdot 10^{-10}\,(\sim 7\cdot 10^{-3}\%)$\\
$K^+K^-$ &\,\,\,\,$7.0\cdot 10^{-10}\,(\sim 0.1\%)$ &\,\,\,\, $6.1\cdot 10^{-10}\,(\sim 0.5\%)$ &\,\,\,\, $3.4\cdot 10^{-11}\,(\sim 5\cdot 10^{-3}\%)$ &\,\,\,\, $2.2\cdot 10^{-11}\,(\sim 0.02\%)$\\
$K^+\pi^-$ &\,\,\,\,$5.9\cdot 10^{-10}\,(\sim 1\%)$ &\,\,\,\, $5.3\cdot 10^{-10}\,(\sim 3\%)$ &\,\,\,\, $1.0\cdot 10^{-10}\,(\sim 0.2\%)$ &\,\,\,\, $7.1\cdot 10^{-11}\,(\sim 0.4\%)$\\
\hline
\end{tabular}
\end{center}
\caption{\small{\it{Branching ratios for: (i) the interference between short distances ($Q_{10}$) and long-distance direct emission (denoted $SD-LD$) and (ii) the pure short-distance interference between $Q_9$ and $Q_{10}$ (denoted $SD-SD$) for the different decay modes. In parenthesis we include the value of the forward-backward asymmetry $A_{FB}$. As reference values for the semileptonic coefficients we have chosen $C_{9,10}= 1$. Results for different values of $C_{9,10}$ can be easily obtained by noting that the first two columns are proportional to $C_{10}$, while the last two are proportional to $C_9C_{10}$. }}}\label{tab3}
\end{table*}
\noindent ${\cal{H}}_{\mu}$ admits a decomposition in terms of form factors ${\cal{F}}_{1,2,3}$ akin to the long-distance one in Eq.~(\ref{param}). As in the long-distance analysis of Section~\ref{sec:II}, the dominant contribution comes from the near-resonant region. Thus,
\begin{align}
\langle h_1h_2|J_{\mu}^{{\bar{u}}c}|D^0\rangle=\langle h_1h_2|{\cal{H}}|V\rangle\frac{1}{P_V(p^2)}\langle V|J_{\mu}^{{\bar{u}}c}|D^0\rangle
\end{align}
Following the steps of Section~\ref{sec:II} and the expressions given in Appendix~\ref{sec:VII}, the form factors can be shown to be:
\begin{align}
{\cal{F}}_1^V(q,p)&=b^V\frac{{\hat{g}}_2(q^2)q\cdot (p_1-p_2)+{\hat{g}}_3(q^2)}{P_V(p^2)}\nonumber\\
{\cal{F}}_2^V(q,p)&=b^V\frac{{\hat{g}}_2(q^2)q\cdot (p_1-p_2)-{\hat{g}}_3(q^2)}{P_V(p^2)}\nonumber\\
{\cal{F}}_3^V(q,p)&=-2b^V\frac{{\hat{g}}_4(q^2)}{P_V(p^2)}
\end{align}
where
\begin{align}
{\hat{g}}_2(q^2)&=-\frac{2i}{m_D+m_{\rho}}{\hat{A}}_2(q^2)\nonumber\\
{\hat{g}}_3(q^2)&=i(m_D+m_{\rho}){\hat{A}}_1(q^2)\nonumber\\
{\hat{g}}_4(q^2)&=\frac{2}{m_D+m_{\rho}}{\hat{V}}(q^2)
\end{align}
and we have used that $b_1^V=b_2^V\equiv b^V$. 

The interference with the long-distance (photon-mediated) contribution studied in Section~\ref{sec:II} reads
\begin{align}\label{eq:EMinterf}
&Re[{\cal{M}}_{LD}^*{\cal{M}}_{SD}^{(10)}]\!=\!\frac{2e}{q^2}\sum_{i<j}G_{ij}{\mathrm{Im}}[{\cal{F}}_i{{F}}_j^*\!-\!{\cal{F}}_j{{F}}_i^*]
\end{align}
where $G_{ij}$ are given by~\cite{Cappiello:2011qc}
\begin{align}
G_{12}&=\epsilon_{\mu\nu\lambda\rho}p_1^{\mu}p_2^{\nu}Q^{\lambda}q^{\rho}\nonumber\\
G_{13}&=-\left[p_1\cdot Q(q^2p_1\cdot p_2-p_1\cdot qp_2\cdot q)\right.\nonumber\\
&\left.+p_2\cdot Q((p_1\cdot q)^2-m_{h1}^2q^2)
\right]\nonumber\\
G_{23}&=\left[p_2\cdot Q(q^2p_2\cdot p_1-p_2\cdot qp_1\cdot q)\right.\nonumber\\
&\left.+p_1\cdot Q((p_2\cdot q)^2-m_{h2}^2q^2)
\right]
\end{align}
The term proportional to $G_{12}$ describes the interference between the electric components of long and short distances, while $G_{13}$ and $G_{23}$ collect the electric {\it{vs}} magnetic interference terms. They contribute to the second line in Eq.~(\ref{angular2}). The forward-backward asymmetry in ${\theta_{\ell}}$:
\begin{align}\label{forback}
A_{FB}&=\langle{\mathrm{sgn}}(c_{\ell})\rangle=\frac{1}{\Gamma}\left[\int_0^1\!\!dy\frac{d\Gamma}{dy}-\int_{-1}^0\!\!dy\frac{d\Gamma}{dy}\right]
\end{align}
singles out ${\cal{A}}_6$ and it is sensitive to the electric {\it{vs}} magnetic interference. Due to the tiny value for $C_{10}$ in the standard model discussed above, a nonvanishing $A_{FB}$ is a rather clean test for new physics.

In Fig.~\ref{fig:5} we show the differential electric-magnetic interference as a function $m_{ll}^2$ for the different decay modes. The figure actually shows the absolute value of the interference: since strong phases are accounted for as Breit-Wigner widths, the contribution flips sign at each resonance pole. For comparison, we have included the long-distance background (dashed lines). Notice that the angular asymmetry stays rather constant from the dilepton threshold up until the resonance peak, beyond which it falls off rather steeply. In order to be quantitative, and following the discussion of Section~\ref{newphysics}, we have considered a new physics scenario in which $C_{10}=1$. The expected size of $A_{FB}$ for the different decay modes is summarized in the first two columns of Table~\ref{tab3}. Notice that since $C_{10}$ is flavor-universal, {\it{i.e.}} not subject to the CKM hierarchy, the signal steadily increases from the Cabibbo-allowed to the doubly Cabibbo-suppressed modes, where it can reach the $3\%$ level for the dimuon mode. 

This last observation suggests to consider also the interference between $C_9$ and $C_{10}$ in new physics scenarios where $C_9\sim C_{10}\sim {\cal{O}}(1)$. In this case the CKM hierarchy is absent altogether: both $C_9$ and $C_{10}$ are flavor-blind. The results are summarized in Fig.~\ref{fig:5} and the second column of Table~\ref{tab3}. Analytical expressions can be found in Appendix~\ref{sec:VIII}. Notice, as compared to the previous case, that the gradual dominance as one goes from the Cabibbo-allowed to the doubly Cabibbo-suppressed is more pronounced, as expected. However, we note that even with this huge enhancement in $C_9$ and $C_{10}$, the purely short-distance interference is only appreciable for the doubly Cabibbo-supressed modes. 


\section{Conclusions}\label{sec:V}
The recent evidence of CP violation in $D^0\to h^+h^-$ decays well above the expected standard model prediction makes the study of D decays a priority in the search of new physics. In this paper we have provided the first detailed analysis of the rare 4-body decays $D^0\to h_1^+h_2^-{\ell}^+{\ell}^-$, $(h=\pi,K; {\ell}=e,\mu)$ in the standard model. We have studied the dominant long-distance contributions (Bremsstrahlung and hadronic effects) in the $(m_{ll}^2,m_{hh}^2)$ Dalitz plots and the total branching ratios, which turn out to be substantially larger than previously estimated. Both the Dalitz plots and the associated branching ratios should be useful tools in view of the upcoming analyses of LHCb and BESIII.
  
For the Cabibbo-allowed, singly Cabibbo-suppressed and doubly Cabibbo-suppressed modes one finds 
\begin{align}\label{branchingratios}
Br[D^0\to K^-\pi^+{\ell}^+{\ell}^-]\sim 10^{-5}\nonumber\\
Br[D^0\to \pi^+\pi^-{\ell}^+{\ell}^-]\sim 10^{-6}\nonumber\\
Br[D^0\to K^+K^-{\ell}^+{\ell}^-]\sim 10^{-7}\nonumber\\
Br[D^0\to K^+\pi^-{\ell}^+{\ell}^-]\sim 10^{-8}
\end{align}
where an important contribution comes from the near-resonant processes $D^0\to V_h^*V_{\ell}^*$, with subsequent decays $V_h^*\to h_1h_2$ and $V_{\ell}^*\to \gamma^*\to {\ell}^+{\ell}^-$. Assuming factorization, lowest meson dominance and strong phases coming mainly from the resonance widths, the different form factors involved can be determined from experimental input. Our results for the hadronic contribution are in agreement with Ref.~\cite{Burdman:2001tf}, which is quite a nontrivial agreement, given that the assumptions and methods going into both analyses are rather different. Finally, we want to emphasize that our results comply with Low's theorem, {\it{i.e.}}, the main contribution in the low dilepton invariant mass region comes from the Bremsstrahlung. 

After having determined the standard-model contribution to $D^0\to h_1^+h_2^-{\ell}^+{\ell}^-$ we have also explored signals for new physics detection. In particular, we have shown that two angular asymmetries, namely the T-odd diplane asymmetry and the forward-backward dilepton asymmetry can provide direct tests of new physics due to tiny standard model backgrounds. Motivated by new physics scenarios proposed to explain $\Delta a_{CP}$ (supersymmetric and Z-enhanced models), we estimate the size of the short-distance parameters $C_9$ and $C_{10}$ compatible with $\Delta a_{CP}$ and flavor constraints. We show that new physics effects in $D^0\to h_1^+h_2^-{\ell}^+{\ell}^-$ can generically reach the $\%$ level.


\section*{Acknowledgments}
We want to thank Gino Isidori and S\'ebastien Descotes-Genon for useful discussions. O.~C.~wants to thank the University of Naples for very pleasant stays during the different stages of this work. G.~D'A. is grateful to the organizers of the Workshop '{\it{Implications of LHCb measurements and future prospects}}' (CERN, April 2012), where this work started. L.~C.~ and G.~D'A.~ are supported in part by the EU under Contract MTRN-CT-2006-035482 (FLAVIAnet) and by MUIR, Italy, under Project 2005-023102. O.~C.~is supported in part by the DFG cluster of excellence 'Origin and Structure of the Universe'. This work was performed in the context of the ERC Advanced Grant project 'FLAVOUR' (267104).\\


\appendix
\section{Kinematics}\label{sec:VI}
We define $p=p_1+p_2$ and $q=k_++k_-$ as the momenta of the dihadron and dilepton pairs, respectively. Then one can obtain for the phase space
\begin{align}
d\Phi &=\frac{1}{4m_D^2}(2\pi)^5\int d p^2 \int dq^2 \sqrt{\lambda_D}\Phi_{h}\Phi_{\ell}
\end{align}  
where 
\begin{align}
\Phi_{h}&=\frac{1}{(2\pi)^5}\frac{1}{8p^2}\sqrt{\lambda_h}\int
d\cos\theta_h\nonumber\\
\Phi_{\ell}&=\frac{1}{(2\pi)^6}\frac{1}{8}\sqrt{1-\frac{4m_\ell^2}{q^2}}\int
d\phi\int d\cos\theta_\ell
\end{align}
Above we have defined $\lambda_h\equiv \lambda(p^2,m_{h1}^2,m_{h2}^2)$ and $\lambda_D\equiv \lambda(p^2,m_D^2,q^2)$  where $\lambda(a,b,c)=a^2+b^2+c^2-2ab-2ac-2bc$. The angular variables are defined as in Ref.~\cite{Cabibbo:1965zz}: if ${\bf{p_1}}$ is the $h_1$ momentum in the dihadron CM system; ${\bf{k_+}}$ the ${\ell}^+$ momentum in the dilepton CM system; ${\bf{{\hat{n}}}}$ the direction of the dihadron system as seen from the $D^0$ rest frame; and ${\bf{p_1^{\perp}}}$ and ${\bf{k_+^{\perp}}}$ the components of ${\bf{p_1}}$ and ${\bf{k_+}}$ perpendicular to ${\bf{{\hat{n}}}}$, then 
\begin{equation}
\cos\theta_{h}=\frac{\mathbf{\hat{n}}\cdot\mathbf{p_1}}{|\mathbf{p_1}|};\,\,\,\,\, \cos\theta_{\ell}=-\frac{\mathbf{\hat{n}}\cdot\mathbf{k_+}}{|\mathbf{k_+}|};\,\,\,\,\,
\cos\phi=\frac{\mathbf{p_1^{\perp}}\cdot\mathbf{k_+^{\perp}}}{|\mathbf{p_1^{\perp}}||\mathbf{k_+^{\perp}}|}
\end{equation}
Intuitively, $\theta_{\ell}$ is the angle between the ${\ell}^+$ momentum and the dihadron system as measured from the dilepton CM while $\phi$ is the angle between the dihadron and dilepton planes.

The final result for the phase space therefore reads
\begin{align}\label{ps31}
d^5\Phi&=\frac{1}{2^{14}\pi^6m_D^2}\frac{1}{p^2}\sqrt{1-\frac{4m_\ell^2}{q^2}}\sqrt{\lambda_D\lambda_h}\nonumber\\
&\times  dp^2 d q^2 d\cos\theta_h d\cos\theta_\ell d\phi
\end{align}
where the range of the kinematical variables is
\begin{align}\label{Kinevar}
4m_{\ell}^2 \, &\le \, q^2\, \le \,(m_D-(m_{h1}+m_{h2}))^2\nonumber\\
(m_{h1}+m_{h2})^2 \, &\le \, p^2 \, \le \,(m_D-\sqrt{q^2})^2\nonumber\\
0\,&\le \, (\theta_h,~\theta_\ell) \, \le \, \pi\nonumber\\
0 \, &\le \, \phi \, \le \, 2\pi
\end{align}
The relevant kinematic products can be expressed as:
\begin{align}
p_1\cdot p_2 &=\frac{1}{2}(p^2-m_{h1}^2-m^2_{h2})\nonumber\\
q\cdot p_{1,2}&=\frac{1}{4}(m_D^2-p^2-q^2)\zeta_{\pm}\mp\frac{1}{4p^2}
\sqrt{\lambda_h\lambda_D}\cos\theta_h\nonumber\\
Q\cdot p_{1,2}&=\beta_\ell\left[-\sqrt{\lambda_D}\frac{\zeta_{\pm}}{4}\cos\theta_\ell
\pm\frac{m_D^2-p^2-q^2}{4}\right.\nonumber\\
&\!\!\!\!\!\!\!\!\!\!\!\!\left.\times\frac{\sqrt{\lambda_h}}{p^{2}} \cos\theta_h
\cos\theta_\ell\mp\frac{\sqrt{\lambda_h}}{2p^2}\sqrt{q^2 p^2}
\sin\theta_h\sin\theta_\ell\cos\phi\right]\nonumber
\end{align}
\begin{align}
\epsilon_{\mu\nu\lambda\rho}p_1^{\mu}p_2^{\nu}q^{\lambda}Q^{\rho}&=\sqrt{\frac{q^2}{p^2}}\frac{\sqrt{\lambda_h\lambda_D}}{4}\beta_{\ell}\sin\theta_{h}\sin\theta_{\ell}\sin\phi
\end{align}
where
\begin{align}
\beta_{\ell}=\sqrt{1-\frac{4m_{\ell}^2}{q^2}};\quad \zeta_{\pm}=\displaystyle\left(1\pm\frac{\chi^2}{p^2}\right)
\end{align}
and $\chi^2=m_{h1}^2-m_{h2}^2$.


\section{Hadronic parameterization}\label{sec:VII}
The hadronic matrix element for the weak vertex can be written, using vector meson dominance and assuming factorization, as
\begin{align}\label{expres}
&\langle V(p)\gamma^*(q)| J^{\mu}_{ij} J_{\mu}^{{\bar{u}}c} | D^0(P)\rangle=\sum_{V^{\prime}}\langle\gamma|{\cal{H}}_{V\gamma}|V^{\prime}\rangle\frac{1}{P_{V^{\prime}}(q^2)}\nonumber\\
&\times \Big\{f_Vm_V\epsilon^{\mu*}_{(V)}\langle V^{\prime}|J^{{\bar{u}}c}_{\mu}|D^0\rangle+f_{V^{\prime}}m_{V^{\prime}}\epsilon^{\mu*}_{(V^{\prime})}\langle V|J^{{\bar{u}}c}_{\mu}|D^0\rangle\Big\}
\end{align}
The most general parametrization for the remaining matrix elements is
\begin{align}
\langle V(p,\epsilon)|&J_{\mu}^{{\bar{u}}c}|D^0(P)\rangle=D_1(k^2)p_{+\mu}+D_2(k^2)k_{\mu}\nonumber\\
&+D_3(k^2)\epsilon_{\mu}^*+iD_4(k^2)\epsilon_{\mu\nu\lambda\rho}p_+^{\nu}k^{\lambda}\epsilon^{\rho*}
\end{align}
with $p_+=P+p$ and $k=P-p$. A similar expression holds for $\langle V^{\prime}|J^{{\bar{u}}c}_{\mu}|D^0\rangle$. The divergence of the current is proportional to the difference of quark masses, which can be parametrized as the squared mass difference of the hadrons. In other words,
\begin{align}
k^{\mu}\langle V(p,\epsilon)|&J_{\mu}^{{\bar{u}}c}|D^0(P)\rangle\sim (m_D^2-m_V^2)k\cdot \epsilon^*
\end{align}
which is satisfied if
\begin{align}
D_1(k^2)&=A_1(k^2)k\cdot \epsilon^*\nonumber\\
D_2(k^2)&=A_2(k^2)k\cdot \epsilon^*\frac{m_D^2-m_V^2}{k^2}\nonumber\\
D_3(k^2)&=A_3(k^2)(m_D^2-m_V^2)
\end{align}
In particular, the second equation above implies that $A_2(0)=0$, such that the divergence is avoided. If one now shifts $A_2\to A_2-A_3-A_1$ one obtains
\begin{align}\label{general}
&\langle V(p)| J_{\mu}^{{\bar{u}}c} | D^0(P)\rangle = A_{1}(k^2)k\cdot\varepsilon^*\nonumber\\
&\times\left[ p_+^{\mu}-\frac{(m_D^2-m_V^2)}{k^2}k^{\mu}\right]+A_2(k^2)k\cdot \varepsilon^* \frac{(m_D^2-m_V^2)}{k^2}k^{\mu}\nonumber\\
&+A_3(k^2)(m_D^2-m_V^2)\left[\varepsilon^{*\mu}- \frac{k\cdot \varepsilon^*}{k^2}k^{\mu}\right]\nonumber\\
&+iD_4(k^2)\varepsilon^{\mu\nu\lambda\rho}p_{+\nu}k_{\lambda}\varepsilon_{\rho}^*
\end{align}
which can be straightforwardly compared with the parametrization of Ref.~\cite{Wirbel:1985ji}. The form factors are related as
\begin{align}
A_1(k^2)&=-i(m_D+m_V)^{-1}{\hat{A}}_2(k^2)\nonumber\\
A_3(k^2)&=i(m_D-m_V)^{-1}{\hat{A}}_1(k^2)\nonumber\\
D_4(k^2)&=i(m_D+m_V)^{-1}{\hat{V}}(k^2)
\end{align}
The hatted form factors are determined in Ref.~\cite{Wirbel:1985ji} in the nearest pole approximation (vector meson resonance) as
\begin{align}
{\hat{V}}(k^2)=\frac{h_{V1}m_{V1}^2}{m_{V1}^2-k^2}&~;\qquad {\hat{A}}_1(k^2)=\frac{h_{A1}m_{A1}^2}{m_{A1}^2-k^2};\nonumber\\
&\!\!\!\!\!\!\!\!\!{\hat{A}}_2(k^2)=\frac{h_{A2}m_{A2}^2}{m_{A2}^2-k^2} 
\end{align}

On the other hand, one has that
\begin{align}\label{ours}
&\langle V(p)\gamma^*(q)| J^{\mu}_{ij} J_{\mu}^{{\bar{u}}c} | D^0(P)\rangle=T^{\mu\nu}(p,k)\epsilon_{\mu}^*(p)\epsilon_{\nu}^*(q)\nonumber\\
&=t_2^V(q\cdot\epsilon_V^*)(p\cdot\epsilon_{\gamma}^*)+t_3^V(\epsilon_{V}^*\cdot \epsilon_{\gamma }^*)+t_4^V\varepsilon^{\mu\nu\lambda\rho}\epsilon_{\mu}^{V*}\epsilon_{\nu}^{\gamma*}p_{\lambda}q_{\rho}
\end{align}
Comparison between Eqs.~(\ref{expres}) and (\ref{ours}) yields the expressions appearing in Eq.~(\ref{formfactors1}) in the main text.

Regarding the strong vertex, the most general parametrization is
\begin{align}
B^{\mu}(p_+,p_-)&= B_1(p_+^2) p_+^{\mu}+B_2(p_+^2)p_-^{\mu}
\end{align}
where $p_{\pm}=p_1{\pm}p_2$. Since
\begin{align}
p_{+\mu}B^{\mu}(p_+,p_-)\sim (m_{h1}^2-m_{h2}^2)
\end{align}
this entails that
\begin{align}
B_1(p_+^2)&=\kappa_1(p_+^2)\frac{m_{h1}^2-m_{h2}^2}{p_+^2}\nonumber\\
B_2(p_+^2)&=\kappa_2(p_+^2)
\end{align}
where $\kappa_1(0)=0$ to smoothen the divergence out. Therefore,
\begin{align}
B^{\mu}(p_1,p_2)&=(\kappa_2+\kappa_1\chi^2)p_1^{\mu}-(\kappa_2-\kappa_1\chi^2)p_2^{\mu}\nonumber\\
&\equiv b_1p_{1}^{\mu}+b_2p_{2}^{\mu}
\end{align}
where $\chi^2\equiv(m_{h1}^2-m_{h2}^2)p_+^{-2}$. In the equal dihadron mass case, $b_1=b_2$, in agreement with charge conjugation invariance. In the general case one finds that $b_1\neq b_2$. However, since $b_i\sim {\cal{O}}($GeV$)$, the mass correction amounts at most to an ${\cal{O}}(1\%)$ correction and can be safely neglected. 


\section{Differential decay widths}\label{sec:VIII}
In this appendix we will provide analytical expressions for the angular-integrated differential decay widths for the generic case $D^0\to h_1^+h_2^-{\ell}^+{\ell}^-$, {\it{i.e.}}, for $m_{h1}\neq m_{h2}$. The singly Cabibbo-suppressed case, {\it{i.e.}}, $m_{h1}=m_{h2}$, can then be worked out as a particular case. We start with  
\begin{equation}
d\Gamma=\frac{1}{2m_D}\sum_{spins} |{\cal M}|^2d\Phi
\end{equation}
where the differential phase space $d\Phi$ is given in Appendix~\ref{sec:VI} and 
\begin{align}
{\cal{M}}={\cal{M}}_{LD}+{\cal{M}}_{SD}^{(9)}+{\cal{M}}_{SD}^{(10)}
\end{align} 
whose definitions can be found in the main text. It is useful to define the following functions:
\begin{align}
h_{+-}(q^2,p^2)&=(m_D^2+q^2-p^2)\nonumber\\
h_{--}(q^2,p^2)&=(m_D^2-q^2-p^2)\nonumber\\
h_{-+}(q^2,p^2)&=(m_D^2-q^2+p^2)\nonumber\\
h_{++}(q^2,p^2)&=(m_D^2+q^2+p^2)
\end{align}
and the mass combinations
\begin{align}
\chi^2&=m_{h1}^2-m_{h2}^2\nonumber\\
{\bar{m}}^2&=m_{h1}^2+m_{h2}^2
\end{align}
For the long-distance contributions with symmetric angular integration one finds
\begin{widetext}
\begin{align}
\frac{d^2\Gamma}{dq^2dp^2}\Bigg|_{Br.}&=\zeta_B\Bigg[\frac{q^2-4m_{h1}^2}{(\chi^2 h_{--}+p^2h_{+-})^2-\lambda_h\lambda_D}+
\frac{q^2-4m_{h2}^2}{(\chi^2 h_{--}-p^2h_{+-})^2-\lambda_h\lambda_D}+\frac{2({\bar{m}}^2-p^2)+q^2}{2\sqrt{\lambda_h\lambda_D}p^2h_{+-}}\nonumber\\
&\times \log{\frac{\chi^4h_{--}^2-(p^2h_{+-}-\sqrt{\lambda_h\lambda_D})^2}{\chi^4h_{--}^2-(p^2h_{+-}+\sqrt{\lambda_h\lambda_D})^2}}\Bigg]\\
\frac{d^2\Gamma}{dq^2dp^2}\Bigg|_{El.}&=\frac{\zeta_E}{|P_V|^2}\Bigg[\bigg\{p^2(p^2-2{\bar{m}}^2)\lambda_D^2+4\chi^4(\lambda_D+3q^2p^2)\lambda_D\bigg\}|a_{21}^V|^2+\bigg\{4h_{--}(h_{+-}^2-4m_D^2q^2)(\lambda_h+3\chi^4)\bigg\}\nonumber\\
&\times Re\bigg[a_{21}^Va_{31}^{V*}\bigg]+4\bigg\{p^2(p^2-2{\bar{m}}^2)(\lambda_D+12q^2p^2)+4\chi^4(\lambda_D+3q^2p^2)\bigg\}|a_{31}^V|^2\Bigg]
\end{align}
\begin{align}
\frac{d^2\Gamma}{dq^2dp^2}\Bigg|_{Mag.}&=\zeta_M\frac{|a_4^V|^2}{|P_V|^2}\\
\frac{d^2\Gamma}{dq^2dp^2}\Bigg|_{Br.-El.}&=\zeta_{EB}\Bigg[\left(8h_{-+}Re\left[\frac{a_{21}^V}{P_V}\right]+16Re\left[\frac{a_{31}^V}{P_V}\right]\right)-
\frac{p^2}{\sqrt{\lambda_h\lambda_D}}\sum_j\log{\frac{(-1)^{j}p^2h_{+-}-\chi^2 h_{--}-\sqrt{\lambda_h\lambda_D}}{(-1)^{j}p^2h_{+-}-\chi^2 h_{--}+\sqrt{\lambda_h\lambda_D}}}\nonumber\\
&\times\left(2h_{+-}\bigg[(-1)^{j}2\chi^2-h_{-+}\bigg]Re\left[\frac{a_{21}^V}{P_V}\right]+4\bigg[(-1)^{j}2\chi^2-h_{++}+8m_{hj}^2\bigg]Re\left[\frac{a_{31}^V}{P_V}\right]\right)\Bigg]
\end{align}
\end{widetext}
where
\begin{align}
\zeta_B&=\frac{\alpha^2}{48\pi^3m_D^3}(2m_l^2+q^2)\beta_{\ell}\frac{\sqrt{\lambda_h\lambda_D}p^2}{q^4}|{\cal{M}}_D|^2\nonumber\\
\zeta_E&=\frac{\alpha}{36864\pi^4m_D^3}(2m_l^2+q^2)\beta_{\ell}\frac{\sqrt{\lambda_h\lambda_D}}{q^6p^6}\nonumber\\
\zeta_M&=\frac{\alpha}{18432\pi^4m_D^3}(2m_l^2+q^2)\beta_{\ell}\frac{(\lambda_h\lambda_D)^{3/2}}{q^4p^4}\nonumber\\
\zeta_{EB}&=-\frac{\alpha^{3/2}}{1536\pi^{7/2}m_D^3}(2m_l^2+q^2)\beta_{\ell}\frac{\sqrt{\lambda_h\lambda_D}}{q^4p^2}|{\cal{M}}_D|
\end{align} 
and we have defined ${\cal{M}}_D\equiv{\cal{M}}_{(D\to h_1h_2)}$. We have included for completeness the Bremsstrahlung {\it{vs.}} electric interference, even though its effect is extremely suppressed. Expressions for the mode-dependent form factors $a_{21}^V$, $a_{31}^V$ and $a_4^V$ can be found in the main text.

For the short-distance contribution entering $A_{\phi}$ one finds that 
\begin{widetext}
\begin{align}
\frac{d^2\Gamma^*}{dq^2dp^2}\Bigg|_{El.-Mag.}&=\frac{\zeta_{EM}}{|P_V(p^2)|^2}\sin\delta_W Re\bigg[a_{31}^Va_4^{V*}\bigg]\label{C10}\\
\frac{d^2\Gamma^*}{dq^2dp^2}\Bigg|_{Br.-Mag.}&=\zeta_{BM}\sin{\delta_S} Re\left[\frac{a_4^V}{P_V}\right]\left[\frac{h_{+-}}{2}\sqrt{\lambda_h\lambda_D}+\bigg\{m_{h1}^2(m_D^2h_{--}-p^2h_{+-})+q^2(\chi^4+m_D^2p^2)-m_{h2}^2q^2h_{-+}\bigg\}\right.\nonumber\\
&\left.\times \log{\frac{p^2h_{+-}+\chi^2 h_{--}-\sqrt{\lambda_h\lambda_D}}{p^2h_{+-}+\chi^2 h_{--}+\sqrt{\lambda_h\lambda_D}}}\right]+\bigg(m_{h1}\leftrightarrow m_{h2},\chi^2\leftrightarrow-\chi^2\bigg)
\end{align}
\end{widetext}
where the asterisk indicates that the integration over $\phi$ is done as in Eq.~(\ref{angularasymm}). $\delta_W$ and $\delta_S$ are, respectively, weak and strong phases and
\begin{align}
\zeta_{EM}&=-\frac{\alpha}{2304\pi^5m_D^3}\beta_{\ell}^3\frac{\lambda_h^{3/2}\lambda_D}{q^2p^4}\nonumber\\
\zeta_{BM}&=-\frac{\alpha^{3/2}}{96\pi^{9/2}m_D^3}\beta_{\ell}^3\frac{\sqrt{\lambda_h}}{q^2}|{\cal{M}}_D|
\end{align}
Again for completeness we have included the Bremsstrahlung {\it{vs}} magnetic interference term, despite being strongly suppressed.

Finally, we list the expression for the contributions entering the forward-backward asymmetry $A_{FB}$, namely the terms $Re[{\cal{M}}_{LD}^*{\cal{M}}_{SD}^{(10)}]$ and $Re[{\cal{M}}_{SD}^{(9)*}{\cal{M}}_{SD}^{(10)}]$: 
\begin{align}
\frac{d^2\Gamma_{FB}}{dq^2dp^2}\Bigg|_{SD-LD}&=\frac{\zeta_{FB}^{(1)}}{|P_V|^2}Re\bigg[a_{31}^V{\hat{a}}_4^{V*}+{\hat{a}}_{31}^Va_4^{V*}\bigg]\label{C13}\\
\frac{d^2\Gamma_{FB}}{dq^2dp^2}\Bigg|_{SD-SD}&=\frac{\zeta_{FB}^{(2)}}{|P_V|^2}Re\bigg[{\hat{a}}_{31}^V{\hat{a}}_4^{V*}\bigg]\label{C14}
\end{align}
The subindex $FB$ is a reminder that the integration over $\theta_{\ell}$ is done using the prescription of Eq.~(\ref{forback}), and
\begin{align}
\zeta_{FB}^{(1)}&=\frac{\alpha^{1/2}\xi_{10}}{6144\pi^{9/2}m_D^3}\beta_{\ell}^2\frac{\lambda_h^{3/2}\lambda_D}{p^4}\nonumber\\
\zeta_{FB}^{(2)}&=\frac{\xi_9\xi_{10}}{6144\pi^5m_D^3}\beta_{\ell}^2\frac{\lambda_h^{3/2}\lambda_Dq^2}{p^4}
\end{align}
and we have used the short-hand notation
\begin{align}
{\hat{a}}_{21}^V(q^2)&=-ib^V{\hat{g}}_2(q^2)\nonumber\\
{\hat{a}}_{31}^V(q^2)&=-ib^V{\hat{g}}_3(q^2)\nonumber\\
{\hat{a}}_{4}^V(q^2)&=-2b^V{\hat{g}}_4(q^2)
\end{align}
Expressions for the short-distance coefficients $\xi_{9,10}$ and the short-distance form factors ${\hat{g}}_i(q^2)$ can be found in the main text.



\begin{thebibliography}{999}

\bibitem{Aaij:2011in} 
  R.~Aaij {\it et al.}  [LHCb Collaboration],
  Phys.\ Rev.\ Lett.\  {\bf 108}, 111602 (2012)
  [arXiv:1112.0938 [hep-ex]].

\bibitem{Collaboration:2012qw} 
  T.~Aaltonen {\it et al.}  [CDF Collaboration],
  Phys.\ Rev.\ Lett.\  {\bf 109}, 111801 (2012)
  [arXiv:1207.2158 [hep-ex]].
  
\bibitem{Aubert:2007if} 
  B.~Aubert {\it et al.}  [BaBar Collaboration],
  Phys.\ Rev.\ Lett.\  {\bf 100}, 061803 (2008)
  [arXiv:0709.2715 [hep-ex]].

\bibitem{Staric:2008rx} 
  M.~Staric {\it et al.}  [Belle Collaboration],
  Phys.\ Lett.\ B {\bf 670}, 190 (2008)
  [arXiv:0807.0148 [hep-ex]].

\bibitem{Amhis:2012bh} 
  Y.~Amhis {\it et al.}  [Heavy Flavor Averaging Group Collaboration],
  arXiv:1207.1158 [hep-ex].
  
\bibitem{Bhattacharya:2012ah} 
  B.~Bhattacharya, M.~Gronau and J.~L.~Rosner,
  Phys.\ Rev.\ D {\bf 85}, 054014 (2012)
  [arXiv:1201.2351 [hep-ph]].
  
\bibitem{Feldmann:2012js} 
  T.~Feldmann, S.~Nandi and A.~Soni,
  JHEP {\bf 1206}, 007 (2012)
  [arXiv:1202.3795 [hep-ph]].
  
\bibitem{Brod:2012ud} 
  J.~Brod, Y.~Grossman, A.~L.~Kagan and J.~Zupan,
  arXiv:1203.6659 [hep-ph].

\bibitem{Isidori:2011qw} 
  G.~Isidori, J.~F.~Kamenik, Z.~Ligeti and G.~Perez,
  Phys.\ Lett.\ B {\bf 711}, 46 (2012)
  [arXiv:1111.4987 [hep-ph]].
  
\bibitem{Franco:2012ck} 
  E.~Franco, S.~Mishima and L.~Silvestrini,
  JHEP {\bf 1205}, 140 (2012)
  [arXiv:1203.3131 [hep-ph]].

\bibitem{Aaij:2013bra} 
  RAaij {\it et al.}  [ LHCb Collaboration],
  arXiv:1303.2614 [hep-ex].

\bibitem{Tilburg}
  J.~van Tilburg, {\it{'New Results on CP violation in the charm sector'}}. CERN-LHC seminar, 12 March 2013.    
\bibitem{Isidori:2012yx} 
  G.~Isidori and J.~F.~Kamenik,
  arXiv:1205.3164 [hep-ph].

\bibitem{Bediaga:2012py} 
  R.~Aaij {\it et al.}  [LHCb Collaboration],
  arXiv:1208.3355 [hep-ex].
See also the talk by B.~Viaud in the Workshop '{\it{Implications of LHCb measurements and future prospects}}' (CERN, April 2012).   
 	   
\bibitem{Aitala:2000kk} 
  E.~M.~Aitala {\it et al.}  [E791 Collaboration],
  Phys.\ Rev.\ Lett.\  {\bf 86}, 3969 (2001)
  [hep-ex/0011077].

\bibitem{Freyberger:1996it} 
  A.~Freyberger {\it et al.}  [CLEO Collaboration],
  Phys.\ Rev.\ Lett.\  {\bf 76}, 3065 (1996)
  [Erratum-ibid.\  {\bf 77}, 2147 (1996)].

\bibitem{BESIII}
http://bes.ihep.ac.cn/bes3/index.html  

\bibitem{Bigi:2011em} 
  I.~I.~Bigi and A.~Paul,
  JHEP {\bf 1203}, 021 (2012)
  [arXiv:1110.2862 [hep-ph]].

\bibitem{D'Ambrosio:1992bf} 
  G.~D'Ambrosio, M.~Miragliuolo and P.~Santorelli,
  In *Maiani, L. (ed.) et al.: The DAPHNE physics handbook, vol. 1* 231-279
  
\bibitem{Low:1958sn} 
  F.~E.~Low,
  Phys.\ Rev.\  {\bf 110}, 974 (1958).

\bibitem{PDG}
	J.~Beringer {\it{et al.}} (Particle Data Group), J.\ Phys.\ D {\bf{86}}, 010001 (2012)    

\bibitem{Cabibbo:1965zz} 
  N.~Cabibbo and A.~Maksymowicz,
  Phys.\ Rev.\  {\bf 137}, B438 (1965)
  [Erratum-ibid.\  {\bf 168}, 1926 (1968)].

\bibitem{Pais:1968zz}
  A.~Pais and S.~B.~Treiman,
  Phys.\ Rev.\  {\bf 168}, 1858 (1968).

\bibitem{Altmannshofer:2008dz} 
  W.~Altmannshofer, P.~Ball, A.~Bharucha, A.~J.~Buras, D.~M.~Straub and M.~Wick,
  JHEP {\bf 0901}, 019 (2009)
  [arXiv:0811.1214 [hep-ph]].

\bibitem{Elwood:1995xv} 
  J.~K.~Elwood, M.~B.~Wise and M.~J.~Savage,
  Phys.\ Rev.\ D {\bf 52}, 5095 (1995)
  [Erratum-ibid.\ D {\bf 53}, 2855 (1996)]
  [hep-ph/9504288].

\bibitem{Pichl:2000ab} 
  H.~Pichl,
  Eur.\ Phys.\ J.\ C {\bf 20}, 371 (2001)
  [hep-ph/0010284].

\bibitem{Cappiello:2011qc} 
  L.~Cappiello, O.~Cata, G.~D'Ambrosio and D.~-N.~Gao,
  Eur.\ Phys.\ J.\ C {\bf 72}, 1872 (2012)
  [arXiv:1112.5184 [hep-ph]].

\bibitem{Burdman:1995te} 
  G.~Burdman, E.~Golowich, J.~L.~Hewett and S.~Pakvasa,
  Phys.\ Rev.\ D {\bf 52}, 6383 (1995)
  [hep-ph/9502329].

\bibitem{Bauer:1986bm} 
  M.~Bauer, B.~Stech and M.~Wirbel,
  Z.\ Phys.\ C {\bf 34}, 103 (1987).

\bibitem{Wirbel:1985ji} 
  M.~Wirbel, B.~Stech and M.~Bauer,
  Z.\ Phys.\ C {\bf 29}, 637 (1985).

\bibitem{Ball:2006eu} 
  P.~Ball, G.~W.~Jones and R.~Zwicky,
  Phys.\ Rev.\ D {\bf 75}, 054004 (2007)
  [hep-ph/0612081].

\bibitem{Link:2004uk} 
  J.~M.~Link {\it et al.}  [FOCUS Collaboration],
  Phys.\ Lett.\ B {\bf 607}, 67 (2005)
  [hep-ex/0410067].

\bibitem{Liu:2012bn} 
  C.~Liu,
  arXiv:1207.1171 [hep-ex].

\bibitem{ElHassanElAaoud:1999nx} 
  El Hassan El Aaoud and A.~N.~Kamal,
  Phys.\ Rev.\ D {\bf 59}, 114013 (1999)
  [hep-ph/9910350].
  
\bibitem{Fajfer:1998rz} 
  S.~Fajfer, S.~Prelovsek and P.~Singer,
  Phys.\ Rev.\ D {\bf 58}, 094038 (1998)
  [hep-ph/9805461].

\bibitem{Fajfer:2005ke} 
  S.~Fajfer and S.~Prelovsek,
  Phys.\ Rev.\ D {\bf 73}, 054026 (2006)
  [hep-ph/0511048].
 
\bibitem{Burdman:2001tf} 
  G.~Burdman, E.~Golowich, J.~L.~Hewett and S.~Pakvasa,
  Phys.\ Rev.\ D {\bf 66}, 014009 (2002)
  [hep-ph/0112235].

\bibitem{Buchalla:1995vs} 
  G.~Buchalla, A.~J.~Buras and M.~E.~Lautenbacher,
  Rev.\ Mod.\ Phys.\  {\bf 68}, 1125 (1996)
  [hep-ph/9512380].

\bibitem{Inami:1980fz} 
  T.~Inami and C.~S.~Lim,
  Prog.\ Theor.\ Phys.\  {\bf 65}, 297 (1981)
  [Erratum-ibid.\  {\bf 65}, 1772 (1981)].

\bibitem{Greub:1996wn} 
  C.~Greub, T.~Hurth, M.~Misiak and D.~Wyler,
  Phys.\ Lett.\ B {\bf 382}, 415 (1996)
  [hep-ph/9603417].

\bibitem{Shifman:1976de} 
  M.~A.~Shifman, A.~I.~Vainshtein and V.~I.~Zakharov,
  Phys.\ Rev.\ D {\bf 18}, 2583 (1978)
  [Erratum-ibid.\ D {\bf 19}, 2815 (1979)].

\bibitem{Bertolini:1986th} 
  S.~Bertolini, F.~Borzumati and A.~Masiero,
  Phys.\ Rev.\ Lett.\  {\bf 59}, 180 (1987).

\bibitem{Riazuddin:1993pn} 
  Riazuddin, N.~Paver and F.~Simeoni,
  Phys.\ Lett.\ B {\bf 316}, 397 (1993)
  [hep-ph/9308328].

\bibitem{Dib:1990gr} 
  C.~O.~Dib and R.~D.~Peccei,
  Phys.\ Lett.\ B {\bf 249}, 325 (1990).

\bibitem{Fajfer:2002gp} 
  S.~Fajfer, P.~Singer and J.~Zupan,
  Eur.\ Phys.\ J.\ C {\bf 27}, 201 (2003)
  [hep-ph/0209250].

\bibitem{Giudice:2012qq} 
  G.~F.~Giudice, G.~Isidori and P.~Paradisi,
  JHEP {\bf 1204}, 060 (2012)
  [arXiv:1201.6204 [hep-ph]].

\bibitem{Paul:2011ar} 
  A.~Paul, I.~I.~Bigi and S.~Recksiegel,
  Phys.\ Rev.\ D {\bf 83}, 114006 (2011)
  [arXiv:1101.6053 [hep-ph]].
  
\bibitem{Sehgal:1992wm} 
  L.~M.~Sehgal and M.~Wanninger,
  Phys.\ Rev.\ D {\bf 46}, 1035 (1992)
  [Erratum-ibid.\ D {\bf 46}, 5209 (1992)].

\end{thebibliography}
\end{document}